\documentclass{article}
\PassOptionsToPackage{numbers, compress}{natbib}
\usepackage[preprint]{neurips_2020}
\usepackage[utf8]{inputenc}
\usepackage{xcolor}
\usepackage{acro}
\usepackage{comment}
\usepackage{xparse}
\usepackage{graphicx}
\usepackage{amsmath}
\usepackage{color,soul}
\usepackage{authblk}
\usepackage{xltabular}
\usepackage[toc,page]{appendix}
\usepackage{multirow}
\usepackage{float}
\usepackage[T1]{fontenc}    % use 8-bit T1 fonts
\usepackage{hyperref}       % hyperlinks
\usepackage{url}            % simple URL typesetting
\usepackage{booktabs}       % professional-quality tables
\usepackage{amsfonts}       % blackboard math symbols
\usepackage{nicefrac}       % compact symbols for 1/2, etc.
\usepackage{microtype}      % microtypography
\usepackage{lineno}
\def\eg{\emph{e.\,g.}} 
\def\ie{\emph{i.\,e.}} 
 
\def\etc{\emph{etc.}}

\newcommand{\new}[1]{\textcolor{black}{#1}}

\definecolor{LANCET_panel}{RGB}{246,232,231}

\NewDocumentCommand{\restrict}{smm}{% #2 is a length, #3 is the text
  \setbox0=\vbox{\hsize=#2\relax\raggedright#3\par
    \setbox2=\lastbox\unskip\unpenalty
    \loop
      \setbox4=\box2
      \setbox2=\lastbox\unskip\unpenalty
    \unless\ifvoid2
    \repeat
    \global\setbox1=\box4
  }% end of the \vbox
  \fbox{\IfBooleanTF{#1}{\unhbox}{\box}1}%
}

\DeclareAcronym{CNN}{
short=CNN,
long=Convolution Neural Network,
foreign-plural={}
}
\DeclareAcronym{AI}{
short=AI,
long=Artificial Intelligence,
foreign-plural={}
}
\DeclareAcronym{CAM}{
short=CAM,
long=Class Activation Map,
foreign-plural={}
}
\DeclareAcronym{PRISMA}{
short=PRISMA,
long=Preferred Reporting Items for Systematic reviews and Meta-Analyses,
foreign-plural={}
}
\DeclareAcronym{BI-RADS}{
short=BI-RADS,
long=Breast Imaging Reporting and Data System,
foreign-plural={}
}
\DeclareAcronym{LI-RADS}{
short=LI-RADS,
long=Liver Imaging Reporting and Data System,
foreign-plural={}
}
\DeclareAcronym{HCC}{
short=HCC,
long=Hepatocellular carcinoma,
foreign-plural={}
}
\DeclareAcronym{EDV}{
short=EDV,
long=end-diastolic volume,
foreign-plural={}
}
\DeclareAcronym{GLCM}{
short=GLCM,
long=gray-level co-occurrence matrix,
foreign-plural={}
}
\DeclareAcronym{CT}{
short=CT,
long=computed tomography,
foreign-plural={}
}
\DeclareAcronym{MRI}{
short=MRI,
long=Magnetic Resonance Imaging,
foreign-plural={}
}
\DeclareAcronym{PET}{
short=PET,
long=positron emission tomography,
foreign-plural={}
}
\DeclareAcronym{OCT}{
short=OCT,
long=Optical Coherence Tomography,
foreign-plural={}
}
\DeclareAcronym{WSI}{
short=WSI,
long=Whole Slide Image,
foreign-plural={}
}
\DeclareAcronym{DTI}{
short=DTI,
long=Diffusion Tensor Image,
foreign-plural={}
}
\DeclareAcronym{AD}{
short=AD,
long=Alzheimer's Disease,
foreign-plural={}
}
\DeclareAcronym{GCN}{
short=GCN,
long=Graphical Convolution Network,
foreign-plural={}
}
\DeclareAcronym{CS}{
short=CS,
long=Computer Science,
foreign-plural={}
}
\DeclareAcronym{ML}{
short=ML,
long=Machine Learning,
foreign-plural={}
}
\DeclareAcronym{HCI}{
short=HCI,
long=Human-Computer Interaction,
foreign-plural={}
}
\DeclareAcronym{CXR}{
short=CXR,
long=Chest X-ray,
foreign-plural={}
}
\DeclareAcronym{COPD}{
short=COPD,
long=Chronic Obstructive Pulmonary Disease,
foreign-plural={}
}
\DeclareAcronym{AUC}{
short=AUC,
long=Area under the ROC Curve,
foreign-plural={}
}
\DeclareAcronym{XAI}{
short=XAI,
long=Explainable Artificial Intelligence,
foreign-plural={}
} 

\makeatletter
\newcommand{\printfnsymbol}[1]{%
  \textsuperscript{\@fnsymbol{#1}}%
}
\makeatother
\def\@fnsymbol#1{\ensuremath{\ifcase#1\or *\or \dagger\or \ddagger\or
   \mathsection\or \mathparagraph\or \|\or **\or \dagger\dagger
   \or \ddagger\ddagger \else\@ctrerr\fi}}

\UseRawInputEncoding

\title{Explainable Medical Imaging AI Needs Human-Centered Design: Guidelines and Evidence from a Systematic Review} % right one

\begin{document}
\author[1*]{Haomin Chen}
\author[1*]{Catalina Gomez}
\author[1]{Chien-Ming Huang}
\author[1**]{Mathias Unberath}
\affil[1]{Department of Computer Science, Johns Hopkins University}
\affil[*]{\textit{Co-first author}}
\affil[**]{\textit{Corresponding author: unberath@jhu.edu}}
\affil[ ]{\textit {\{hchen135,cgomezc1,chienming.huang,unberath\}@jhu.edu}}

\maketitle
% published justification
\begin{abstract}
Transparency in \ac{ML} attempts to reveal the working mechanisms of complex models. 
From a human-centered design perspective, transparency is not a property of the \ac{ML} model but an affordance, i.\,e., a relationship between algorithm and user. Thus, prototyping and user evaluations are critical to attaining solutions that afford transparency. Following human-centered design principles in highly specialized and high stakes domains, such as medical image analysis, is challenging due to the limited access to end users. This dilemma is further exacerbated by the high knowledge imbalance between \ac{ML} designers and end users. To investigate the state of transparent \ac{ML} in medical image analysis, we conducted a systematic review of the literature by screening titles, abstracts, and keywords of all available records from January 2012 through July 2021 in PubMed, EMBASE, and Compendex databases. We identified 2508 records and 68 articles met the inclusion criteria. We found that current techniques in transparent \ac{ML} are dominated by computational feasibility, and barely consider end users, during neither development nor evaluation. Despite the considerable difference between the roles and knowledge of \ac{ML} developers and clinical stakeholders, no study reported formative user research to inform the design and development of transparent \ac{ML} models. Only a few studies validated transparency claims through empirical user evaluations. These shortcomings put contemporary research on transparent \ac{ML} at risk of being incomprehensible to users, and thus, clinically irrelevant. To alleviate these shortcomings in forthcoming research while acknowledging the challenges of human-centered design in healthcare, we introduce the \emph{INTRPRT guideline}, a systematic design directive for transparent \ac{ML} systems in medical image analysis. The \emph{INTRPRT guideline} suggests human-centered design principles and highly recommends formative user research as the first step of transparent model design to understand user needs and domain requirements. It further provides alternatives to empirical user evaluation to support transparent \ac{ML} design choices to facilitate the adoption of human-centered design principles. Following these guidelines and a human-centered design process increases the likelihood that the algorithms afford transparency and enables stakeholders to capitalize on the benefits of transparent \ac{ML}.
\end{abstract}

\acresetall
\begin{figure}[H]
\center
\includegraphics[width=1\textwidth]{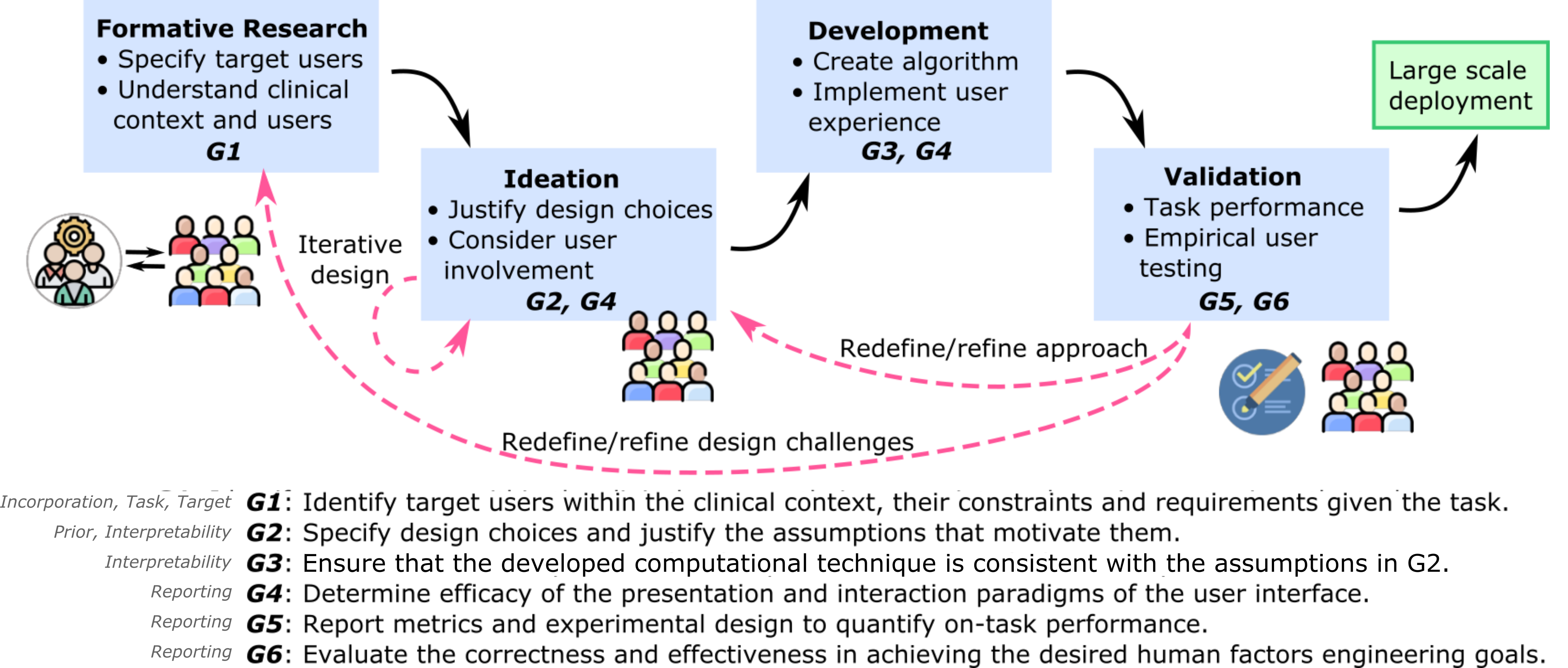}
\caption{Schematic representation of the \emph{INTRPRT guideline} within the main stages of a human-centered design process. The blue boxes demonstrate the process from understanding end users and their context to the validation of the developed system, which ultimately might result in large scale deployment. The guidelines are summarized below and are located within the design phases based on the aspects pertinent to each one \new{and the corresponding themes of each guideline are listed on the left}. Opportunities for iterative design are illustrated with the dashed arrows. }
\label{fig:main_figure_guideline}
\end{figure}

\section*{Introduction}
% \subsection{Machine learning for healthcare with human-machine teaming}
There have been considerable research thrusts to develop \ac{ML} models in the healthcare domain that assist clinical stakeholders~\cite{topol2019high}. However, translating these \ac{ML} models from the bench to the bedside to support clinical stakeholders during routine care brings substantial challenges, among other reasons, because of the high stakes involved in most decisions that impact human lives. When stakeholders interact with \ac{ML} tools to reach decisions, they may be persuaded to follow \ac{ML}'s recommendations that may be incorrect or promote unintended biases against vulnerable populations, all of which can have dreadful consequences~\cite{obermeyer2019dissecting}. 
These circumstances motivate the need for trustworthy \ac{ML} systems in healthcare and have sparked efforts to specify the different requirements that \ac{ML} algorithms should fulfill. 
Most of these recent efforts focus on achieving a certain on-task performance requirement but neglect that for assisted decision making not \ac{ML} system performance alone, but human-\ac{ML} team performance is the most pertinent to patient outcome. How to achieve adequate human-machine teaming performance, however, is debated. While some argue that rigorous algorithmic validation, e.\,g., similar to the evaluation of drugs, tests, or devices, demonstrates safe and reliable operation and may thus be sufficient for successful human-machine teaming~\cite{ghassemi2021false,mccoy2022believing}, others reason that transparency in an \ac{ML} model, e.\,g., by revealing its working mechanisms \new{and presenting a proper interface}, is necessary to invoke user trust and achieve the desired human-machine teaming performance~\cite{vellido2020importance,char2020identifying,holzinger2019causability}. 
\new{The growing interest and convergence of recent works on the importance and need of transparency have stressed that not addressing the opacity of \ac{ML} techniques might hinder their adoption of in healthcare, limiting the potential positive impacts~\cite{markus2021role,salahuddin2022transparency,vellido2020importance,banegas2021towards,ploug2020four,amann2020explainability}.
The inability to make the decision making process transparent might affect the misuse and disuse of \ac{ML} models in the clinical domain, as the utility of the model might be limited if it does not reveal the reasoning process, limitations, and biases \cite{salahuddin2022transparency}.}
We believe that this dichotomy is artificial in that, first, rigorous validation and transparency are not mutually exclusive, and second, both approaches augment an \ac{ML} model with additional information in hopes to justify (in other words, make transparent) the recommendation's validity which is hypothesized to achieve certain human-factors engineering goals such as understandability, reliability, trust and etc. 
However, as we will highlight in detail through a systematic review, current approaches that aim at advancing human factors goals of \ac{ML} systems rely on developers' intuition rather than considering whether these mechanisms affect users' experience with the system and their ability to act on \ac{ML} model's outputs.

% \subsection{A human-centered design approach to transparency in machine learning}
Designing \ac{ML} algorithms that are transparent is fundamentally different from merely designing \ac{ML} algorithms. The desire for transparency adds a layer of complexity that is not necessarily computational. Rather, it involves human factors, namely the users to whom the \ac{ML} algorithm should be transparent. As a consequence, transparency of an algorithm is not a property of the algorithm but a relationship between the transparent \ac{ML} algorithm and the user processing the information. Such relationship can be understood as an \emph{affordance}, a concept that is commonly employed when designing effective \acp{HCI}~\cite{norman1999affordance}, and we argue that transparency in \ac{ML} algorithms should be viewed as such. 
There are several consequences from this definition:
\begin{itemize}
    \item Developing transparent \ac{ML} algorithms is not purely computational.
    \item Specific design choices on the mechanisms to achieve explanations or interpretations may be suitable for one user group, but not for another. 
    \item Creating transparent \ac{ML} systems without prior groundwork to establish that it indeed affords transparency may result in misspent effort.
\end{itemize}

Given the user- and context-dependent nature of transparency, it is essential to understand the target audience and to validate design choices through iterative empirical user studies to ensure that design choices of transparent models are grounded in a deep understanding of the target users and their context. In addition, to maintain a user-centered approach to design from the early stages, rapid prototyping with users provides feedback on the current, low- to high-fidelity embodiment of the system that is going to be built eventually. Involving users early by exposing them to low-fidelity prototypes that mimic final system behavior allows designers to explore multiple alternatives before committing to one pre-determined approach that may not be understandable nor of interest to end users.

However, following a human-centered design approach to build transparent \ac{ML} systems for highly specialized and high stakes domains, such as healthcare, is challenging. The barriers are diverse and include: 1) the high knowledge mismatch between \ac{ML} developers and the varied stakeholders in medicine, including providers, administrators, or patients; 2) availability restrictions or ethical concerns that limit accessibility of potential target users for iterated empirical tests in simulated setups for formative research or validation; 3) challenges inherent to clinical problems, including the complex nature of medical data (\eg, unstructured or high dimensional) and decision making tasks from multiple data sources; and last but not least, 4) the lack of \ac{ML} designers' training in design thinking and human factors engineering.

Starting from the considerations around designing and validating transparent \ac{ML} for healthcare presented above, we investigate the current state of transparent \ac{ML} in medical image analysis, a trailblazing application area for \ac{ML} in healthcare due to the abundance and structure of data. Through a systematic review based on these aspects, we first identify major shortcomings in the design and validation processes of developing transparent \ac{ML} models. These deficiencies include the absence of formative user research, the lack of empirical user studies, and in general, the omission of considering \ac{ML} transparency as contingent on the targeted users and contexts. Together, these shortcomings of contemporary practices in transparent \ac{ML} development put the resulting solutions at substantial risk of being unintelligible to the target users, and consequently, irrelevant. 

This paper aims to encourage model designers to actively consider and work closely with the end users during the design, construction, and validation of \ac{ML} models for medical imaging problems. Acknowledging the barriers to widespread adoption of human-centered design techniques to develop transparent \ac{ML} in healthcare and grounded in our systematic review of the literature, we further propose the \emph{INTRPRT guideline} to help model designers for developing transparent \ac{ML} for medical image analysis step by step. \new{Figure~1 summarizes our guideline within a human-centered design process.} 
The guideline aims at highlighting the need to ground and justify design choices in a solid understanding of the users and their context when adding transparency or other human factors-based goals to \ac{ML} systems for medical image analysis. By raising awareness of the user- and context-dependent nature of transparency, designers should consider a trade-off between efforts to 1) better ground their approaches on user needs and domain requirements and 2) commit to technological development and validation of possibly transparent systems. In this way, the guideline may increase the likelihood for algorithms that advance to the technological development stage to afford transparency, because they are well grounded and justified in user and context understanding. This may mitigate misspent efforts in developing complex systems without prior formative user research, and help designers make accurate claims about transparency and other human factors engineering goals when building and validating the model. To the best of our knowledge, we provide the first guidelines for models that afford transparency and involve end users in the design process for medical image analysis.

\section*{An overview of current trends in transparent machine learning development}
\new{Compared to developing generic \ac{ML} algorithms, designing and validating transparent \ac{ML} algorithms in medical imaging tasks requires consideration of human factors and clinical context. We group these additional considerations into six themes according to the initial review, iteratively defined prior to data extraction and abbreviated to \emph{INTRPRT}; the themes are incorporation (IN), interpretability (IN), target (T), reporting (R), prior (PR), and task (T). {\em Incorporation} refers to the communication and cooperation between designers and end users before and during the construction of the transparent model. Formative user research is one possible strategy that can help designers to understand end users' needs and background knowledge~\cite{cai2019human,xie2020chexplain}, but other approaches exist~\cite{jacobs2021designing}. {\em Interpretability} considers the technicalities of algorithmic realization of a transparent \ac{ML} system. Figure~2 provides illustrative examples of some of these techniques. {\em Target} determines the end users of the transparent \ac{ML} algorithms. {\em Reporting} summarizes all aspects pertaining to the validation of transparent algorithms. This includes task performance evaluation as well as the assessment of technical correctness and human factors of the proposed transparency technique (\eg, intelligibility of the model output, trust, or reliability). {\em Prior} refers to previously published, otherwise public, or empirically established sources of information about target users and their context. This prior evidence can be used to conceptualize and justify design choices around achieving transparency. Finally, \emph{task} specifies the considered medical image analysis task, such as prediction, segmentation, or super resolution, and thus determines the clinical requirements on performance. We emphasize that these themes should not be considered in isolation because they interact with and are relevant to each other. For example, technical feasibility of innovative transparency mechanisms based on the desired task may influence both, the priors that will be considered during development as well as the incorporation of target users to identify and validate alternatives.}

\new{Having identified and refined the themes iteratively after an initial review, we structured} the systematic review according to the six themes. We identify and summarize dominant trends among the 68 included studies aiming to design transparent \ac{ML} for medical image analysis. \new{In the \emph{incorporation} theme,} cross-disciplinary study teams may constitute a first step towards incorporating target users during \ac{ML} design, however, only 33 of the included articles were authored by multidisciplinary clinician-engineering teams. More importantly, no paper introduced formative user research to understand user needs and contextual considerations before model construction, \new{which is reflected in the lack of justifying the \emph{prior} theme. Around half of the selected articles (n=28) chose clinical priors and guidelines as an inspiration for transparent systems. In the \emph{target} theme,} we found that only 30 of the included articles specified end users, and all of these papers were aimed at clinical care providers, a stark imbalance considering the variety of stakeholders. 
\new{In the \emph{task} theme,} prediction tasks were by far the most common application for transparent \ac{ML} algorithm design (57/68). 
\new{In the \emph{Interpretability} theme,} methods relying on clinical guidelines resulted in algorithms that adopted multiple sub-steps of a clinical guideline to build the model and generate outcomes, while methods that were based on computer vision techniques for transparency most commonly relied on post-hoc explanations. 
\new{In the \emph{Reporting} theme,} the methods used for assessing transparency varied with the problem formulation and transparency design, and included human perception, qualitative visualizations, quantitative metrics, and empirical user studies; we note that an evaluation with end users was highly uncommon (only 3 of the 68 included studies). 
\new{However, no paper considered the six themes comprehensively. More importantly, there is no evidence that any papers considered the dependency and interaction between different themes. The reviewed literature further supports that one guideline considering all themes and the interaction between them is highly desired in the medical image analysis community to construct transparent \ac{ML} models following human-centered design practices.}

\section*{INTRPRT guideline} 
\label{sec:guideline}

We distilled a set of guidelines for designing transparent \ac{ML} models according to the interaction and relevancy among the six themes, which is proposed here as \emph{INTRPRT guideline}. The \emph{INTRPRT guideline} provides suggestions for designing and validating transparent \ac{ML} systems in healthcare in hopes to increase the likelihood that the resulting algorithms indeed afford transparency for the designated end users. The guidelines also address the challenges of following a human-centered design approach in the healthcare domain, propose potential solutions, and apply to different kinds of transparency \ac{ML} algorithms. \new{To further illustrate the \emph{INTRPRT guideline}, we introduce a case study (see Supplementary information A).}

\begin{figure}
\center
\includegraphics[width=1\textwidth]{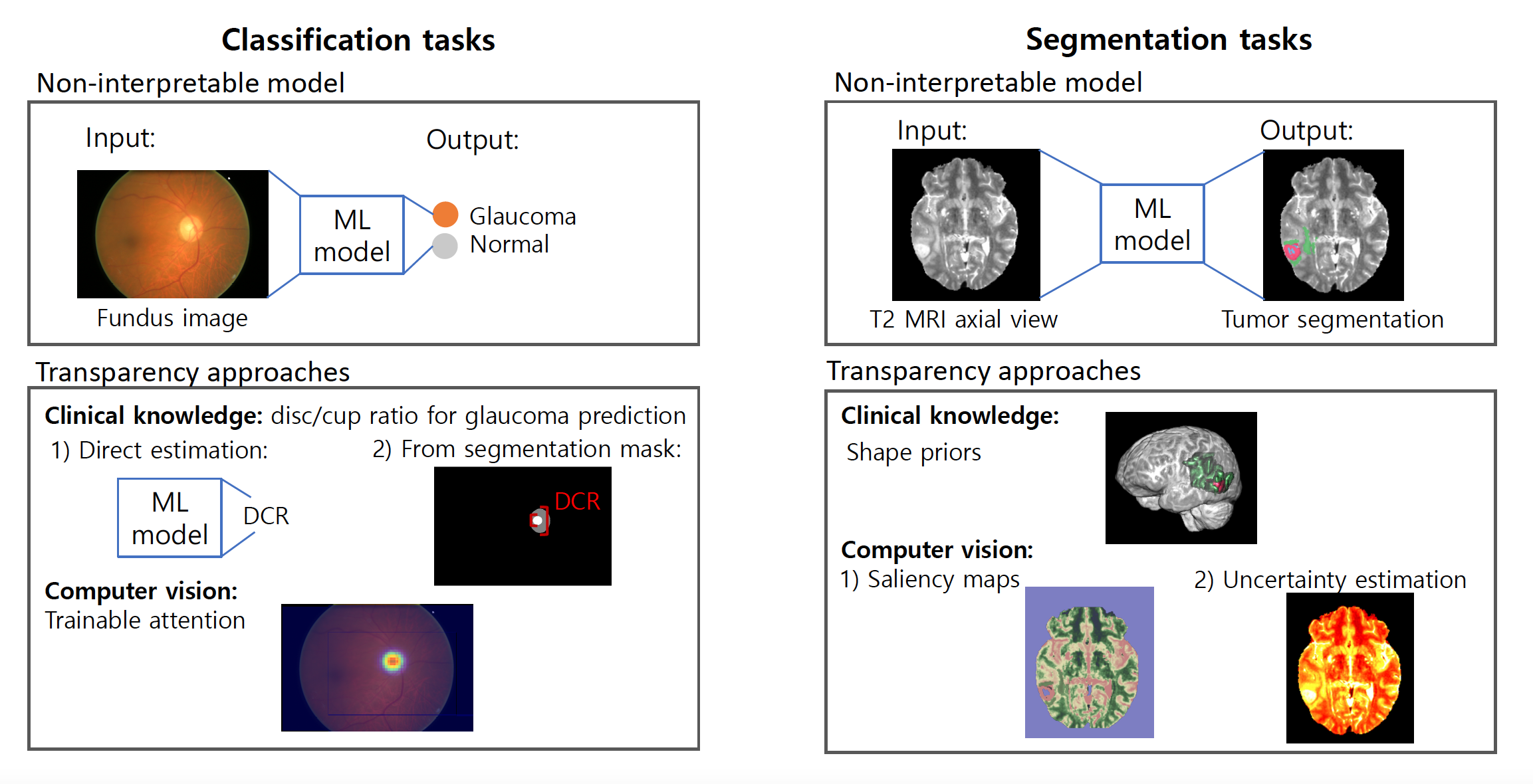}
\caption{Illustrative examples of different techniques used in transparent \ac{ML} systems for classification and segmentation tasks from the systematic review. Images retrieved from the ORIGA~\cite{zhang2010origa} and BraTS2020 datasets~\cite{menze2014multimodal}.}
\label{fig:examples}
\end{figure}

\textbf{Guideline 1: Specify the clinical scenario, constraints, requirements, and end users.} 
The first step to designing any \ac{ML} algorithm for healthcare is to well define the clinical scenario, the constraints the solution will have to abide by, and all hard or soft requirements the algorithm needs to meet for the clinical {\em task} to be addressed adequately (cf.~Figure~3). For \ac{ML} algorithms that do not attempt to be transparent, it is essential but sufficient to assess whether the envisioned \ac{ML} algorithm design will satisfy the clinical constraints and requirements, \eg, an acceptable classification accuracy in allowable processing time. In addition, when designing transparent \ac{ML} algorithms it is equally critical to determine and characterize the end users. It is of particular importance to investigate end user characteristics specifically in the clinical context of the chosen task. This is because, depending on the task, stakeholders have varied interest, prior knowledge, responsibilities, and requirements~\cite{suresh2021beyond}. Deep understanding of the role target users play in the chosen clinical task and their unique needs is critical in determining how to achieve transparency (Guideline 2).

\textbf{Guideline 2: Justify the choice of transparency and determine the level of evidence.}
There exists a wide gap in domain expertise and contextualization between target users and \ac{ML} model designers in most use cases in healthcare. \new{Furthermore, there are multiple ``transparency'' techniques and choices, such as the transparent working mechanism or user-friendly \ac{HCI}.} Simply selecting a ``transparency'' technique, without incorporating and consulting target users puts the resulting \ac{ML} models at risk of not achieving the desired transparency. The human-centered design approach addresses this challenge through iterative empirical studies that over time guide the development and refinement of the technical approach such that, upon completion, the design choices are well justified by empirical target user feedback. This approach may not always be feasible in healthcare due to accessibility and availability barriers of target users. To address this limitation while still enabling technological progress in transparent \ac{ML}, we introduce four distinct levels of evidence. These levels allow designers to classify the level of confidence one may have that the specific design choices will indeed result in a model that affords transparency.

The levels of evidence are based on increasingly thorough approaches to understand the chosen end users in context of the envisioned task:
\begin{itemize}
	\item {\bf Level 0: No evidence.} No dedicated investigations about the end users are performed to develop transparent \ac{ML} systems.
	\item {\bf Level 1: One-way evidence.} Formative user research techniques, such as surveys and diary studies, are only performed once without further feedback from end users about the findings extracted from the research phase, resulting in one-way evidence. 
	Such user research suffers risks of potential bias in concluding about justification of transparency because there is no opportunity for dialog, \ie, designers may ask irrelevant questions or target users may provide non-insightful, potentially biased responses. 
	\item {\bf Level 2: Public evidence.} Public evidence refers to information about target user knowledge, preference, or behavior that is public domain and vetted in a sensible way. Public evidence includes clinical best practice guidelines, Delphi consensus reports, peer-reviewed empirical studies of closely related approaches in large cohorts, or well documented socio-behavioral phenomena. 
	\item {\bf Level 3: Iteratively developed evidence.} Iteratively developed evidence is transparency evidence that is iteratively refined through user feedback where designers and end users communicate with each other throughout method development. The purpose of iteratively validating and refining the current transparency mechanism is to identify any potential bias in the assumptions that motivate the transparency technique while ensuring that it is understandable to end users. 
\end{itemize}
Being actively cognizant of the level of evidence that supports the development enables trading off development efforts between \ac{ML} method development vs. gathering richer evidence in support of the intended developments.

\textbf{Guideline 3: Clarify how the model follows the justification of transparency.} 
This guideline is designed to ensure that the transparency technique used in the \ac{ML} model is indeed consistent with the assumptions made during its justification. 
While complying with this guideline is trivial if the model is developed in a human-centered design approach (Level 3: Iteratively developed evidence), in all other cases designers should be explicit about the intellectual proximity of the developed technical approach to the motivating evidence. 
To this end, after specifying which components of the \ac{ML} model require transparency for users to capitalize on the intended benefits, it is desirable for the method to be as simple as possible so that it can be easily derived from and linked to the justification of transparency. Once confirmed that the envisioned models is indeed consistent with the justification, computational development of the model, including training, refinement, and validation, begins. 

\textbf{Guideline 4: Determine how to communicate with end users.} 
In addition to content (Guidelines 1,2, and 3), seemingly peripheral factors on the presentation of information may play a disproportionate role in the perception of transparency. It is well known that factors like format (\eg, text, images, plots)~\cite{lai2019human}, channel (\eg~graphical interface)~\cite{eiband2018bringing}, and interactivity (\eg~whether users can provide feedback or refine model outputs) can drastically affect users' experience and performance~\cite{wang2021explanations,cheng2019explaining,smith2020no}, and therefore, 
must be aligned with the goal of transparent system development. Clearly, the selection of presentation mode should be incorporated early and supported by some degree of evidence, that emerges naturally when following human-centered design principles but requires justification if not (as posited for transparency in Guideline 2).
Ultimately, users' experience with the systems plays an important role in their willingness to adopt it in a real setup~\cite{cheng2019explaining,bansal2021does}.

\textbf{Guideline 5: Report task performance of the \ac{ML} systems.} 
Similar to ordinary \ac{ML} models, the transparent system must be evaluated quantitatively using reliable targets and appropriate metrics that well reflect the desired performance. In addition, the data used to evaluate the algorithm and its relevance regarding the clinical target task must be specified. Metrics and evaluation protocols should be selected to well determine the models ability in regard to the clinical requirements specified per Guideline 1. Reporting task performance of the algorithm in standalone deployment is important as a baseline for empirical studies in which users may interact and collaborate with the algorithm to complete a task, and team performance (human + \ac{ML} system) metrics can be measured. Such comparisons are relevant to the goal of improving team performance when integrating intelligent systems to assist humans in complex tasks~\cite{bansal2019beyond,bansal2021does}. 
\begin{figure}
\center
\includegraphics[width=1\textwidth]{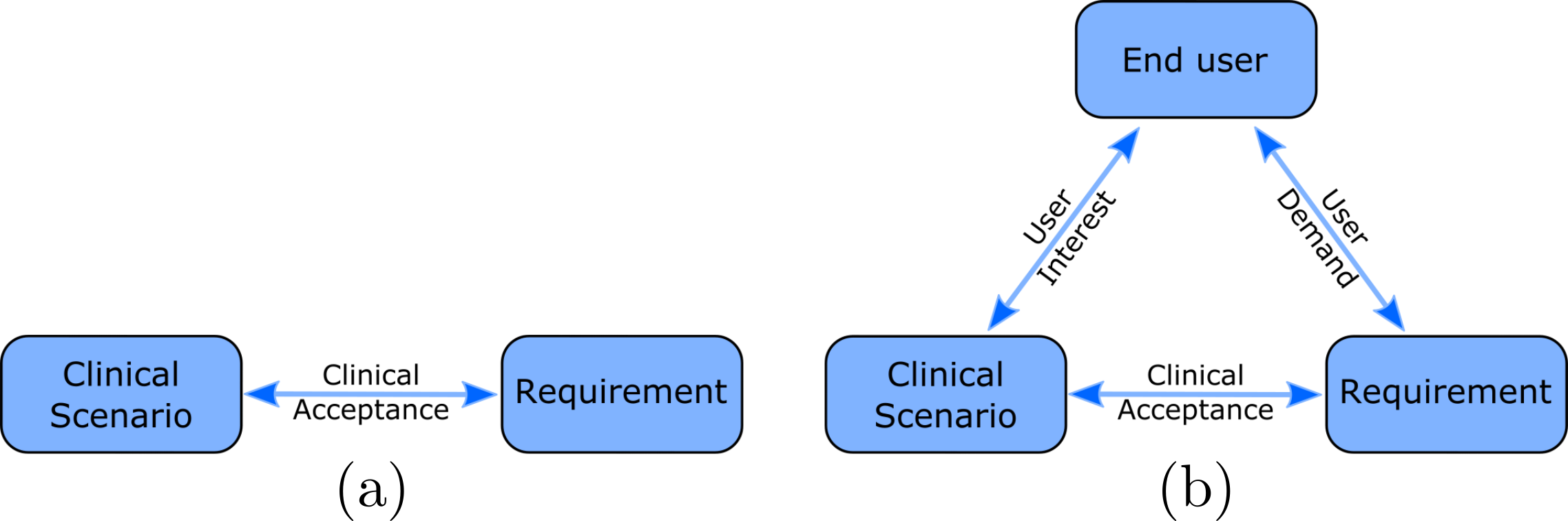} 
\caption{(a) shows the relationship between clinical scenario and requirement in non-transparent \ac{ML} systems, were the system may not be human-facing, and as such, meeting quantitative performance requirements is sufficient.
(b) shows the relationship between clinical scenario, requirement and end users in transparent \ac{ML} systems, as they arise in a human-centered system that seeks to enable users to accomplish a certain task. 
}
\label{fig:G1}
\end{figure}

\textbf{Guideline 6: Assess correctness and human factors of system transparency.}
In addition to task performance, transparent \ac{ML} systems must be evaluated with respect to their transparency claims. This validation is generally necessary even for Level 3 (iteratively developed evidence) justified transparency mechanisms, because user studies in the design phase commonly rely on mock-up prototypes of the envisioned system, and therefore, may have different modes of failure than the final \ac{ML} system. Within this guideline, we distinguish two types of evaluation: 1) Validating the correctness of the transparency technique, which objectively assesses whether the information supplied to achieve transparency is in agreement with the justification. Achieving correctness is particularly important for systems that rely on post-hoc explanations, since explanations may rely on a second model that is distinct from the \ac{ML} algorithm generating recommendations. 2) Validating the effectiveness of transparency in a human-centered approach, to demonstrate that the transparent \ac{ML} system applied to relevant data samples and in cooperation with target users achieves the desired goals. This empirical evaluation determines the efficacy of human factors engineering. The dimensions that are often considered include users' trust~\cite{lai2019human}, reliance, satisfaction~\cite{smith2020no}, mental model formation~\cite{nourani2021anchoring}, and system acceptance.
Reporting of these user studies should include details of the experimental design, participant sample, and techniques to analyze the results.

Transparency of \ac{ML} algorithms for medical image analysis is commonly motivated by the desire of automating complex tasks while retaining a clear interface for human interaction, \eg, to improve trust, avoid over-reliance, or increase acceptance. However, achieving these design goals through transparency requires the development of transparent \ac{ML} algorithms that are intelligible by the envisioned end users. In design thinking, aligning technological developments with user needs is accomplished through user involvement in the design process and iterative user testing, which is largely infeasible in healthcare due to varied barriers to end user involvement. We propose a design and evaluation framework where \ac{ML} designers actively consider end users' needs, knowledge, and requirements, allowing designers to classify the reliability of their understanding of end user needs using four levels of evidence. Explicitly thinking about the confidence one may have in the assumptions about end users that drive transparent \ac{ML} system development may mitigate the risks of developing solutions that are unintelligible to the target users, and therefore, neither achieve the desired human factors engineering goals nor benefit clinical practice. Similarly, quantifying the level of evidence currently available to motivate transparency claims then allows developers to trade-off resources between technical \ac{ML} developments and additional formative research of their target users to ensure that the resulting systems are fit to meet the requirements of the clinical task but foremost the target users. 

\section*{Discussion}
\new{We discuss the \emph{INTRPRT guideline} in the context of the key observations from different themes of our systematic review to identify opportunities to improve the design of transparent ML systems for medical image analysis. Each discussion focuses on one or more themes that we introduced before. Furthermore, the second last subsection presents successful examples of \ac{ML} systems designed with clinical end users and the last subsection also includes the comparison of our guideline and systematic review with existing literature.}

{\bf Importance of formative user research and empirical user testing}
Both formative user research \new{(theme \emph{incorporation} and \emph{prior})} and empirical user testing \new{(theme \emph{reporting})} are critical to ensure that solutions meet user needs \new{(theme \emph{target})}. On the one hand, formative user research helps designers navigate and understand end users' domain practice and needs. On the other hand, empirical user testing assesses whether the designed algorithm indeed achieves the human factors engineering goals, such as affording transparency, promoting trust, or avoiding over-reliance. Additionally, early user involvement in the design process using prototypes of increasing fidelity provides opportunities to review and iterate over design choices. From our systematic review of the literature presented in Methods Section, we find that although most contemporary studies on transparent \ac{ML} formulate human factor engineering goals, no study reported formative user research or empirical testing to inform and validate design choices. We must conclude that contemporary research efforts in medical image analysis have disproportionately prioritized the technological development of algorithmic solutions that alter or augment the predictions of complex \ac{ML} systems with the implicit -- though unfortunately often explicit -- assumption that those changes would achieve transparency. However, because of the substantial knowledge imbalance between \ac{ML} engineers and target users among other reasons detailed above, it is unlikely that, without formative user research or empirical tests, those systems truly afford transparency or achieve the promised human factors engineering goals. While demonstrating the computational feasibility of advanced transparency techniques is certainly of interest, grounding the need for these techniques in solid understanding of the target users should be the first step for most, if not all, such developments. 

{\bf General assessment of transparency}
In addition to user involvement or formative research during the design phase \new{(theme \emph{incorporation} and \emph{prior})}, upon completion of \ac{ML} development, system transparency needs to be empirically validated \new{(theme \emph{reporting})}. During review, we observed that hardly any study reported quantitative empirical user evaluations as part of final method validation, and many of the included articles limited analysis of transparency goals to qualitative analysis by presenting a limited number of illustrative examples, \eg, \new{pixel-attribution} visualizations. 
While such analysis may suggest fidelity of the transparency design to the cause of the prediction in those few select samples, its utility beyond is unclear. In cases where no empirical user evaluation is conducted, neither during conceptualization nor during development, claims around system transparency or human factors are at high risk of being optimistic and should be avoided. 

{\bf Transparent machine learning systems for diverse stakeholders}
\new{The purpose of adding transparency to an \ac{ML} model varies across end users and their context, which we covered in the \emph{target} theme.} 
The current literature on transparent \ac{ML} for medical image analysis focuses heavily on care providers. In fact, all of the included articles that explicitly specify end users targeted clinicians, such as radiologists, pathologists, and physicians. However, clear opportunities for transparent \ac{ML} systems exist for other clinical stakeholders, such as other care team members including nurses or techs, healthcare administrators, insurance providers, or patients. Designing transparent \ac{ML} systems for these stakeholder groups will likely require different approaches, both technological as well in regards to human factors engineering, because these target users are likely to exhibit distinct needs, requirements, prior knowledge, and expectations. In light of recent articles that question the utility of transparency in high stakes clinical decision making tasks~\cite{ghassemi2021false,mccoy2021believing}, driving transparent \ac{ML} development using a ``human factors first'' mindset while expanding target user considerations to more diverse stakeholder groups may increase the likelihood of transparent \ac{ML} having an impact on some aspects of the healthcare system.

{\bf Transparency for tasks with and without human baseline}
\new{Clearly specifying and formulating the medical task that the \ac{ML} system solves is fundamental to determine the assistance that it can provide to clinical practice.} 
Along with the disproportionate consideration of clinicians as end users goes a disproportionate focus on clinical tasks that are routinely performed in current clinical practice (n=60/68) by those target users. One motivation for investigating transparency in such tasks is the existence of clear and systematic clinical workflows and guidelines, \eg, the \ac{BI-RADS} system for mammography, the AO/OTA Tile grading of pelvic fractures, or other easily intelligible covariates associated with outcomes. The availability of such human-defined baselines that are already used for clinical decision making provides immediate Level 2 evidence of transparency for \ac{ML} systems attempting their replication. In addition, it facilitates data collection and annotation, because intermediate outputs that may be required to realize such system are known a priori. 
Conversely, justifying specific attempts at achieving transparency is much more complicated for tasks that do not readily have human-based baselines or clinical best practice guidelines. Some such tasks may already be performed in clinical practice, such as segmentation or super-resolution, the interpretation of which may be ambiguous and result in high variability among observers~\cite{deeley2013segmentation}. Other tasks may be beyond the current human understanding of the underlying mechanisms that enable \ac{ML}-based prediction, \eg, ethnicity prediction from chest X-ray~\cite{banerjee2021reading} or various tasks in digital pathology~\cite{liu2020gene,lu2020deep}. In these scenarios, while it may be possible to derive some justification from the literature, \eg, how target users generally approach tasks of the kind, achieving even Level 2 justification is difficult if not impossible. Empirically validating the envisioned mechanisms for transparency with respect to their ability to afford transparency and achieve the human factors engineering goals is thus paramount when attempting to benefit such tasks.

{\bf Successful examples of \ac{ML} systems designed with clinical end users}
\label{sec:discussion_example}
Early identification and direct communication with end users, \new{as it is emphasized in the \emph{target} and \emph{incorporation}  themes,} allows \ac{ML} designers to bridge the knowledge gap and design for users in highly specialized contexts. 
By following human-centered design and \ac{HCI} practices, previous works have illustrated ways to incorporate end users in the design process of \ac{ML} systems for clinical applications. For instance, target users were consulted in the design of an \ac{ML} tool in an image retrieval system for medical decision making~\cite{cai2019human}, enabling the team to design a system that preserves human agency to guide the search process. Through an iterative design process, functional prototypes of different refinement techniques based on documented user needs were implemented and further validated in a user study. To enable users to explore and understand an \ac{AI} enabled analysis tool for \ac{CXR} images, a user-centered iterative design assessed the utility of potential explanatory information in the \ac{AI} system~\cite{xie2020chexplain}. Users' needs during their introduction to an \ac{AI}-based assistance system for digital pathology were identified through open-ended interviews and a qualitative laboratory study~\cite{cai2019hello}. Iterative co-design processes were followed to identify clinicians' perceptions of \ac{ML} tools for real clinical workflows, \eg, antidepressant treatment decisions~\cite{jacobs2021designing} and phenotype diagnosis in the intensive care unit~\cite{wang2019designing}. Determining the efficacy of envisioned \ac{ML} systems or \ac{ML}-enabled interaction paradigms in empirical user studies before committing resources to their fully-fledged implementation has become common practice in human-centered AI, \eg,~\cite{nourani2020role,buccinca2021trust,cheng2019explaining}, with many studies considering tasks that are related to medical image analysis~\cite{gaube2021ai,xie2020chexplain}. Increasing the acceptance of empirical formative user research as an integral component of human-centered \ac{ML} design for healthcare tasks, including medical image analysis, will be critical in ensuring that the assumptions on which human-centered systems are built hold in the real world.

{\bf Increasing demand for guidelines to build \ac{ML} systems}
\label{sec:discussion_comparison}
Motivated by advances in \ac{AI} technologies and the wide range of applications in which it can be used to assist humans, there are ongoing efforts to guide the design and evaluation of \ac{AI}-infused systems that people can interact with \new{(theme \emph{target})}.
% generic guidelines
Generally applicable design guidelines were compiled and iteratively refined by \ac{HCI} experts to design and evaluate human-\ac{AI} interactions~\cite{amershi2019guidelines}. Although these guidelines are relevant and suitable for a wide range of common \ac{AI}-enabled systems, more nuanced guidelines are desirable for domains where study participants cannot be recruited nor interviewed in abundance.
Similarly, previous attempts to guide the design of effective transparency mechanisms acknowledge that real stakeholders involved should be considered and understood~\cite{suresh2021beyond,liao2020questioning,wang2019designing}. 
Starting from the identification of diverse design goals according to users' needs and their level of expertise on \ac{AI} technology, and a categorization of evaluation measures for \ac{XAI} systems,~\cite{mohseni2021multidisciplinary} addressed the multidisciplinary efforts needed to build such systems. A set of guidelines, summarized in a unified framework, suggests iterative design and evaluation loops to account for both algorithmic and human factors of \ac{XAI} systems.
However, similar to~\cite{amershi2019guidelines}, these guidelines are intended for generic applications, \eg, loan and insurance rate prediction~\cite{chen2019fairness} and personalized advertisements~\cite{datta2014automated}, and do not consider additional challenges, barriers, and limitations when developing algorithms for domains that exhibit users with very specific needs and in highly specific contexts, such as healthcare. 
Other considerations to build interpretable \ac{AI} systems have been identified from a multidisciplinary perspective~\cite{amann2020explainability}. For instance, the approach presented in~\cite{leslie2019understanding} summarized four guidelines that included the application domain, technical implementation, and human-centered requirements in terms of the capabilities of human understanding. 
\new{A requirements list formulated as a ``fact sheet'' was introduced in ~\cite{sokol2020explainability} to characterize and assess explainable systems along five key dimensions: functional, operational, usability, safety and validation. While the five dimensions allow to systematically compare and contrast explainability approaches theoretically and practically, the properties that were included failed to consider where and how to formulate the justification of transparency. Formative user research and validation of the justification of transparency are especially essential in healthcare, where a huge knowledge imbalance exists between \ac{ML} designers and end users of \ac{AI} systems. }
% algorithmic solutions may ultimately impact human lives but

% existing guidelines for medical ML
Considering potential uses of \ac{AI} in clinical setups, \new{there have been efforts to define guidelines for the development and reporting of medical \ac{ML} systems.
For instance, guidelines for clinical trials that involve \ac{AI} were proposed in~\cite{liu2020reporting}, including items such as the description of intended users, how the \ac{AI} intervention was integrated, how the \ac{AI} outputs contributed to decision-making, among others. While specifying these items is also relevant for creating transparent systems, these guidelines do not include requirements in dimensions unique to the transparency of an algorithm, such as its justification and validation.
Guidelines for the initial clinical use of \ac{AI} systems were formulated in~\cite{decide2021decide}, highlighting the importance to assess the actual impact of an algorithm on its users' decisions at an early stage. This recommendation of an early and formative evaluation is aligned with our guideline with respect to formative user research during the initial stages to support design choices for transparency. 
Concerned with the reproducibility and reliability of medical \ac{ML} studies, a set of practical guidelines as a checklist or questions have been collected for authors and reviews to assess the methodological soundness of contributions~\cite{cabitza2021need}, to promote standard reporting practices~\cite{hernandez2020minimar}, and for clinicians to assess algorithm readiness for routine care~\cite{scott2021clinician}. Besides the general reporting items regarding the problem definition, data, model, and validation, these checklists consider the definition of the target user and availability of interpretability information and support for related claims; however, these are questions to be solved once the transparency technique has been incorporated and might lack an appropriate justification and not achieve the intended goals. 
Considering the reason to demand explainability in advance, which is determined by the application domain and target users of the \ac{AI} system, can be used to determine the importance and usefulness of the properties offered by certain explainability techniques. To choose among available explainability techniques, a framework with recommendations regarding mostly technical aspects for researchers was proposed in~\cite{markus2021role}. }

\new{With the trend that \ac{ML} is more popular in clinical decision making tasks due to its performance, recent surveys and systematic reviews have aimed to summarize existing literature to create transparent \ac{ML} in healthcare. However, these surveys failed to consider all the themes proposed in this paper and each aspect of transparent \ac{ML} is reviewed in isolation. More importantly, current reviews mainly focus on the existing transparency techniques and evaluation, ignoring how and where justification of transparency emerges.
For example, a survey categorized research works related to the interpretability of ML in general, and then applied the same categories to interpretability in the medical field~\cite{tjoa2020survey}. In addition to providing an overall perspective of the different interpretable algorithms that are available in the medical field, the survey identified the recurring assumption of having interpretable models without human subject tests, questioning the utility within medical practices and whether \ac{ML} designs consider actual medical needs. 
% Several techniques, such as saliency maps, signal methods and feature extraction techniques are summarized. However, listing techniques without considering the actual clinical tasks and clinical users decreases the likelihood of \ac{ML} models to be indeed transparent in real practices.
In particular, there have been surveys focused uniquely on transparent techniques for medical imaging. The interpretability methods to explain deep learning models were categorized in detail based on technical similarities, along with the progress made on the corresponding evaluation approaches in
\cite{salahuddin2022transparency}. 
Another overview of deep learning-based \ac{XAI} in  medical image analysis is presented in~\cite{van2022explainable}, considering a variety of techniques that were adapted or developed to generate visual, textual, and example-based explanations in the medical domain. Some of the observed trends and remarks in this survey match our perspective and recommendations in the design of transparent methods for medical imaging, including the lack of evaluation as a standard practice, the user-dependent nature of explanations, and the importance of active collaboration with experts to include domain information. 
% summarizing existing ``Interpretability'' techniques for transparent \ac{ML}, such as perturbation analysis and Backpropagation-based approaches.
% TODO: connect 
Instead of proposing a general perspective in a broad range of healthcare problems, some reviews focus on specific topics of medical image analysis. Transparent \ac{ML} for human experts in cancer diagnosis with \ac{AI} is reviewed in~\cite{banegas2021towards} with a focus on 2 aspects: \ac{ML} model characteristics that are important in cancer prediction and treatment; and the application of \ac{ML} in cancer cases. These two aspects are similar to our proposed theme ``Interpretability'' and ``task'', but we summarize the two themes in the general medical image analysis area instead of limiting in cancer study, include more on recent studies (starting from 2012), and focus on more recent \ac{ML} techniques such as \acp{CNN}. Likewise, transparent \ac{ML} in cancer detection is also reviewed in~\cite{gulum2021review} and structured following the same aspects of generic transparent \ac{ML} techniques, such as Local vs. Global and Ad-Hoc vs. Post-Hoc. distinctions} 

The guidelines and systematic review of the state of the field presented here aim at emphasizing the need for formative user research and empirical user studies to firmly establish the validity of assumptions on which human factors engineering goals (including transparency) are based; a natural first step in human-centered AI or \ac{HCI}, but not yet in medical image analysis. As methods for the human-centered development of transparent \ac{ML} for medical image analysis mature, the guidelines presented here may require refinements to better reflect the challenges faced then. At the time of writing, supported by the findings of the systematic review, we believe that the lack of explicit formative research is the largest barrier to capitalizing on the benefits of transparent \ac{ML} in medical image analysis.

{\bf Conclusion} Transparency is an affordance of transparent \ac{ML} systems, \ie, a relationship between models and end users. Therefore, especially in contexts where there exists a high knowledge gap between \ac{ML} developers and the envisioned end users, developing transparent \ac{ML} algorithms without explicitly considering and involving end users may result in products that are unintelligible in the envisioned context and irrelevant in practice.
Efforts to build \ac{ML} systems that afford transparency in the healthcare context should go beyond computational advances, which -- based on the findings of our systematic review -- is not common practice in the context of transparent \ac{ML} for medical image analysis. While many of the approaches claimed transparency or derivative accomplishments in human factors engineering, they did so even without defining target users, engaging in formative user research, or reporting rigorous validation. Consequently, for most of the recently proposed algorithms, it remains unclear whether they truly afford transparency or advance human factors engineering goals.  
We acknowledge that building systems that afford transparency by involving end users in the design process is challenging for medical image analysis and related healthcare tasks. In this context, we propose the \emph{INTRPRT guideline} that emphasize the importance of user and context understanding for transparent \ac{ML} design, but provide alternatives to empirical studies for formative user research. By following these guidelines, \ac{ML} designers must actively consider their end users throughout the entire design process. We hope that these design directives will catalyze forthcoming efforts to build transparent \ac{ML} systems for healthcare that demonstrably achieve the desired human factors engineering goals.

\section*{Methods}
\label{sec:systematic_review}
\subsection*{Search strategy and selection criteria}
The aim of the systematic review is to survey the current state of transparent \ac{ML} methods for medical image analysis. Because \ac{ML} transparency as major research thrust has emerged following the omnipresence of highly complex \ac{ML} models, such as deep \acp{CNN}, we limited our analysis to records that appeared after January 2012, which pre-dates the onset of the ongoing surge of interest in learning-based image processing~\cite{krizhevsky2017imagenet}.

We conducted a systematic literature review in accordance with the \ac{PRISMA} method~\cite{moher2009preferred}. We searched PubMed, EMBASE, and Compendex databases to find articles pertinent to transparent \ac{ML} (including but not limited to explainable and interpretable \ac{ML}) for medical imaging by screening titles, abstracts, and keywords of all available records from January 2012 through July 2021. Details of the search terms and strategy can be found in Supplementary information B.

\subsection*{Study selection}
Following the removal of duplicates (1731 remained), studies were first pre-screened using the title and abstract. Studies that did not describe transparent methods nor medical imaging problems were immediately excluded (217 remained). We then proceeded to full-text review, where each study was examined to determine whether the study presented and evaluated a transparent \ac{ML} technique for medical image analysis. Failure to comply with the described inclusion/exclusion criteria resulted in the study's removal from further consideration. Detailed statistics and a complete description of the pre-screening and full-text review can be found in Supplementary information C and Figure~4. $68$ articles were included for information extraction.

\subsection*{Data extraction strategy}
For the $68$ selected articles that met the inclusion criteria, two authors (H.C. and C.G.) performed detailed data extraction to summarize important information related to the six themes described in INTRPRT Guideline Section. A data extraction template was developed by all authors and is summarized in Supplementary information D. Every one of the $68$ articles was analyzed and coded by both authors independently and one author (H.C.) merged the individual reports into a final consensus document.
Despite our efforts to broadly cover all relevant search terms regarding transparent \ac{ML} in medical imaging, we acknowledge that the list may not be exhaustive. There are vast numbers of articles that have imbued transparency in their methodology, but transparency (or contemporary synonyms thereof, such as explainability or interpretability) is not explicitly mentioned in the title, abstract, or keywords of these articles, and often not even in the body of the text~\cite{rudin2019stop}. This fact makes intractable to identify all articles about transparent \ac{ML} methods. Finally, the review is limited to published manuscripts, long articles and novel approaches. Publication bias may have resulted in the exclusion of works relevant to this review.

\begin{figure}
\center
\includegraphics[scale=0.27]{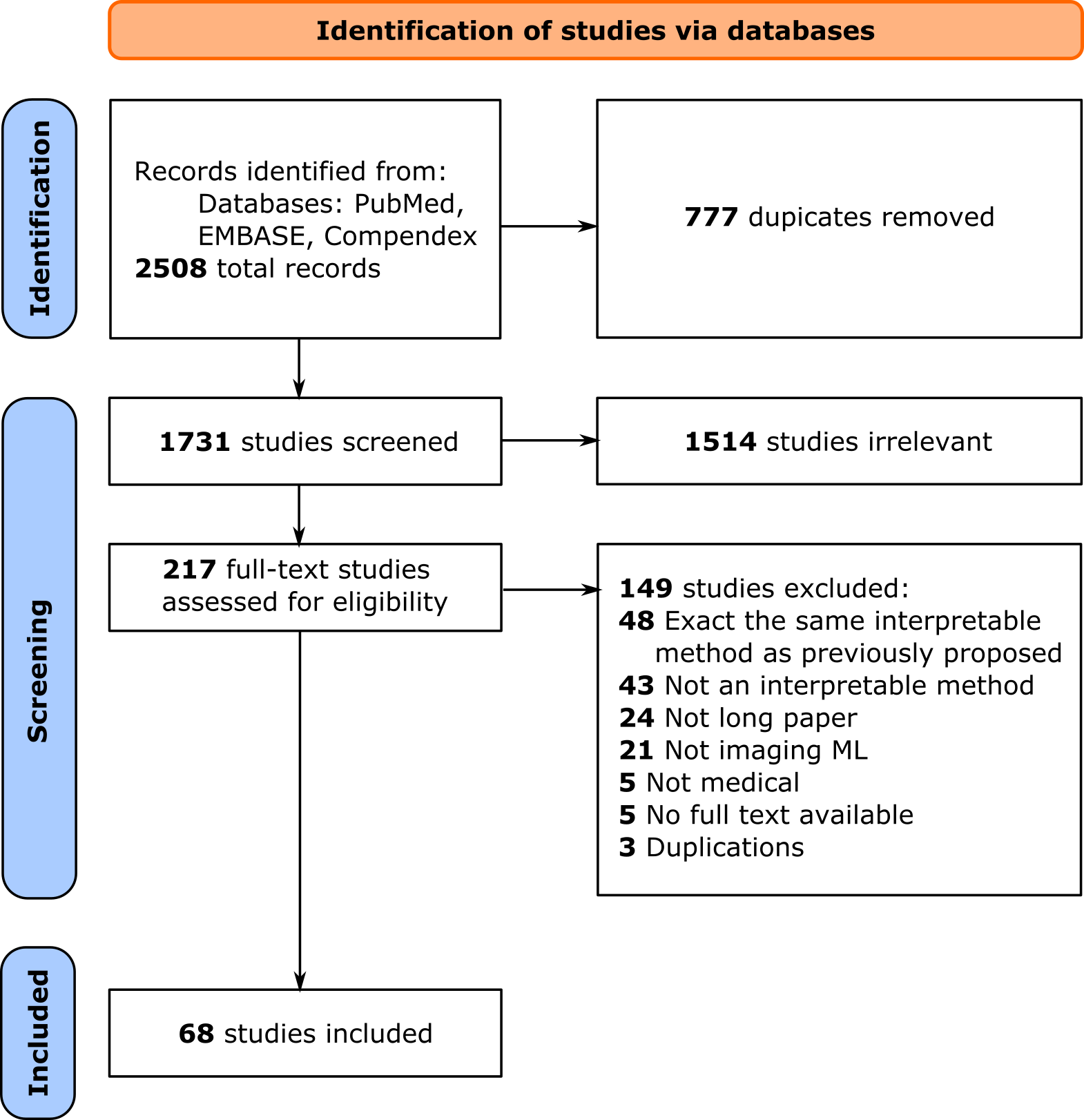}
\caption{PRISMA diagram for transparent \ac{ML} in medical imaging.}
\label{fig:PRISMA}
\end{figure}

\section*{Detailed analysis of findings during systematic review}
The data extraction template for studies included in the systematic review was structured using the six themes of the \emph{INTRPRT guideline}, the adequacy of which was confirmed during data extraction. Therefore, we summarize our findings for each theme and provide details of the extraction results for each article in Tables~{2\&3} in Supplementary information D.

\subsection*{IN: Incorporation}
\label{sec:Incorporation}
A common trend among included studies (n=$33$) was that the presented methods were developed by multidisciplinary clinician-engineering teams, as was evidenced by the incorporation of clinical specialists, such as physicians, radiologists, or pathologists, in the study team and on the author lists. In light of the current bias towards clinicians as end users of transparent \ac{ML} algorithms, this observation suggests that designers may have communicated with a limited subset of the intended end users. However, no formative user research is explicitly described or introduced in these articles to systematically understand the end users before implementing the model.
Further, we found that incorporating clinical experts did not have a considerable impact on whether clinical priors or standard or care guidelines (\ie, Level 2 evidence) were used to build the \ac{ML} system (39\%/44\% articles with/without the incorporation of end users use clinical priors). Regarding the technical approach to provide transparency, the incorporation of medical experts motivated designers to incorporate prior knowledge directly into the model structure and/or inference for medical imaging (73\%/64\% articles with/without the incorporation of end users do not need a second model to generate transparency). 

\subsection*{IN: Interpretability}
\label{sec:iNterpretability}
Transparency of \ac{ML} systems was achieved through various techniques, including attention mechanisms (n=15), use of human-understandable features (n=11), a combination of deep neural networks and transparent traditional \ac{ML} methods (n=7), visualization approaches (n=5), clustering methods (n=4), uncertainty estimation / confidence calibration (n=3), relation analysis between outputs and hand-crafted features (n=3), and other custom techniques (n=20).

The use of an attention mechanism was the most common technique for adding transparency. Attention mechanisms enabled the generation of pixel-attribution methods~\cite{molnar2020interpretable} to visualize pixel-level importance for a specific class of interest~\cite{AbdelMagid2020, Afshar2021, Fan2020, Graziani2020, An2021, He2021, Hou2019, Huang2019, Morvan2020, Saleem2021, Shahamat2020, Singla2018, Sun2020, Xu2021, Yang2020}. In segmentation tasks, where clinically relevant abnormalities and organs are usually of small sizes, features from different resolution levels were aggregated to compute attention and generate more accurate outcomes, as demonstrated in multiple applications, \eg, multi-class segmentation in fetal \acp{MRI}~\cite{Graziani2020} and multiple sclerosis segmentation in~\acp{MRI} \cite{Hou2019}. 
Clinical prior knowledge was also inserted into the attention mechanism to make the whole system more transparent. For instance,~\cite{Shahamat2020} split brain \acp{MRI} into 96 clinically important regions and used a genetic algorithm to calculate the importance of each region to evaluate \ac{AD}. 

Human-understandable features, \eg, hand-crafted low-dimensional features or clinical variables (age, gender, \etc) were frequently used to establish transparent systems. There existed two main ways to use human-understandable features in medical imaging: 1) Extracting hand-crafted features, \eg, morphological and radiomic features, from predicted segmentation masks generated by a non-transparent model~\cite{Diao2021, Dong2021, Giannini2016, Loveymi2020, MacCormick2019, Kunapuli2018, Shen2020, Li2021, Puyol-Anton2020, Wongvibulsin2020} followed by analysis of those hand-crafted features using a separate classification module; 2) Directly predicting human-understandable features together with the main classification and detection task~\cite{Lin2020, Kim2018a, Kim2018, Puyol-Anton2020a, Wang2019a}. In these approaches, all tasks usually shared the same network architecture and parameter weights. 

Instead of explicitly extracting or predicting human-understandable features, other articles further analyzed deep encoded features with human-understandable techniques by following clinical knowledge. Techniques such as decision trees were constructed based on clinical taxonomy for hierarchical learning~\cite{Codella2018,Barata2021, Silva2018,Khaleel2021,Shen2020,Pereira2018,Yan2019}. Rule-based algorithms~\cite{chen2020deep} and regression methods~\cite{Verma2019} were used to promote transparency of the prediction. \cite{Li2020} created a \ac{GCN} based on clinical knowledge to model the correlations among colposcopic images captured around five key time slots during a visual examination.

We also identified various other methods to create transparent systems. These methods can be categorized as visualization-based, feature-based, region importance-based, and architecture modification-based methods. Each approach is discussed in detail below.

Visualization-based methods provide easy-to-understand illustrations by overlaying the original images with additional visual layouts generated from transparency techniques. There existed two main visualization-based methods: 1) Visualizing \new{pixel-attribution maps}: These maps may be generated using gradient-based importance analysis~\cite{Wang2019,Zhao2018}, pixel-level predicted probability~\cite{Folke2021}, or a combination of different levels of feature maps~\cite{Liao2020,Shinde2019}. 2) Latent feature evolution: Encoded features were evolved according to the gradient ascent direction so that the decoded image (\eg, generated with an auto-encoder technique~\cite{ballard1987modular}) gradually change from one class to another \cite{Biffi2018,Couteaux2019}. 

Feature-based methods directly analyze encoded features in an attempt to make the models transparent. Various feature-based transparency method were proposed for transparent learning. \cite{Guo2016,Janik2021,Sari2019} first encoded images to deep features and then clustered samples based on these deep features for prediction or image grouping tasks. Feature importance was also well-studied to identify features that are most relevant for a specific class by feature perturbation \cite{Venugopalan2021,Zhu2019} and gradients \cite{Pirovano2020}. \cite{Hao2020} identified and removed features with less importance for final prediction through feature ranking. 

As an alternative to measure feature contribution, input region importance was also analyzed to reveal sub-region relevance to each prediction class. Image occlusion with blank sub-regions \cite{DeSousa2020,Li2018,Quellec2021} and healthy-looking sub-regions \cite{Uzunova2019} was used to find the most informative and relevant sub-regions for classification and detection tasks. 

Other approaches modified the network architecture according to relevant clinical knowledge to make the whole system transparent. \cite{Graziani2020} pruned the architecture according to the degree of scale invariance at each layer in the network. \cite{Liu2019a} created ten branches with shared weights for ten ultrasound images to mimic the clinical workflow of liver fibrosis stage prediction. \cite{Liu2021} aggregated information from all three views of mammograms and used traditional methods to detect nipple and muscle direction, which was followed by a grid alignment according to the nipple and muscle direction for left and right breasts. \cite{Oktay2018} proposed to learn representations of the underlying anatomy with a convolutional auto-encoder by mapping the predicted and ground truth segmentation maps to a low dimensional representation to regularize the training objective of the segmentation network. 

Some other methods used the training image distribution to achieve transparency in classification. \cite{Peng2019} used similar-looking images (nearest training images in feature space) to classify testing images with majority votes. Causal inference with plug-in clinical prior knowledge also introduced transparency directly to automatic systems \cite{Liu2019,Ren2021,Velikova2013}. Confidence calibration and uncertainty estimation methods were also used to generate additional confidence information for end users \cite{Carneiro2020,Sabol2020,Tanno2021}. 

\subsection*{T: Targets}
\label{sec:Targets}
A striking observation was that none of the selected articles aimed at building transparent systems for users other than care providers. Less than half of the articles explicitly specified clinicians as the intended end users of the system (n=30). From the remaining 38 articles, 17 articles implied that the envisioned end users would be clinicians, while the remaining 21 did not specify the envisioned target users. Articles that were more explicit about their end users were more likely to rely on clinical prior knowledge (Level 2 evidence) in model design. In total, 47\% of articles that specified or implied clinicians as end users implemented clinical prior knowledge in the transparent systems while only 18\% of articles without end user information use clinical prior knowledge.

\subsection*{R: Reporting}
\label{sec:Reporting}
Evaluating different properties of a transparent algorithm besides task-related metrics, especially its performance in regards to achieving the desired human factors engineering goals, complements the assessment of the \ac{ML} model's intended purpose. We identified that the quality of the transparency component is currently being evaluated through four main approaches. 
The first one involves metrics based on human perception, such as the mean opinion score introduced in \cite{Oktay2018} to capture two expert participants' rating of the model's outcome quality and similarity to the ground truth on a 5-point scale. Using two study participants, pathologists' feedback was also requested in \cite{Pirovano2020} to assess their agreement with patch-based visualizations that display features relevant for normal and abnormal tissue. The level of agreement was not formally quantified, but reported as a qualitative description. Similarly, one study participant was involved in a qualitative assessment of explanations quality in \cite{Puyol-Anton2020a,Hao2020}. These evaluations are different from empirical user studies as they are limited to a few individuals and were mostly used to subjectively confirm the correctness of the transparent component.

% quantitative approaches
The second approach attempted to quantify the quality of explanations for a specific purpose (functionally-grounded evaluation~\cite{doshi2017towards}). For instance, some articles evaluated the localization ability of post-hoc explanations by defining an auxiliary task, such as detection~\cite{Khaleel2021,Fan2020}  or segmentation~\cite{Shinde2019,Uzunova2019,Codella2018,Huang2019} of anatomical structures related to the main task. They then contrasted relevant regions identified by the model with ground truth annotations. These quantitative measures (dice score, precision, recall) allowed for further comparisons with traditional explanations methods. Similarly,~\cite{Peng2019} defined a multi-task learning framework for image classification and retrieval, evaluating retrieval precision and providing a confidence score based on the retrieved neighbors as an attempt to check the learned embedding space. Capturing relevant features consistent with human intuition was proposed in~\cite{Zhu2019} by measuring the fraction of reference features recovered, which were defined according to a guideline. Overall, the evaluation of explanations through auxiliary tasks required additional manual efforts to get the necessary ground truth annotations. 

% metrics related to the explanations themselves Silva2018, Khaleel2021, Zhao2018, Tanno2021, Liu2019
Properties of the explanation itself were also quantified as their usefulness to identify risky and safe predictions at a voxel-level for the main task by thresholding on their predictive uncertainty values~\cite{Tanno2021}. Other properties of explanations, such as their correctness (accuracy of rules), completeness (fraction of the training set covered) and compactness (size in bytes) were measured in~\cite{Silva2018}. A measure related to completeness was defined in~\cite{Khaleel2021} and aimed to capture the proportion of training images represented by the learned visual concepts, in addition to two other metrics: the inter- and intra-class diversity and the faithfulness of explanations computed by perturbing relevant patches and measuring the drop in classification confidence. Other articles followed a similar approach to validate relevant pixels or features identified with a transparent method; for example, in \cite{Saleem2021} a deletion curve was constructed by plotting the dice score vs. the percentage of pixels removed and \cite{AbdelMagid2020} defined a recall rate when the model proposes certain number of informative channels. 
\cite{Zhao2018} proposed to evaluate the consistency of visualization results and the outputs of a \ac{CNN} by computing the \textit{L1} error between predicted class scores and explanation \new{pixel-attribution maps}. 
In summary, while the methods grouped in this theme are capable of evaluating how well a method aligns with it's intended mechanism of transparency, they fall short of capturing any human factors-related aspects of transparency design.

The third, and most common approach,
involved a qualitative validation of the transparent systems (n=40) by showing \new{pixel-attribution} visualizations overlaid with the input image or rankings of feature relevance, along with narrative observations on how these visualizations may relate to the main task. These qualitative narratives might include comparisons with other visualization techniques in terms of the highlighted regions or the granularity/level of details. Furthermore, following a retrospective analysis, the consistency between the identified relevant areas/features and prior clinical knowledge in a specific task was a common discussion item in 37\% of all the articles (n=25); refer to articles \cite{Li2018,Liu2019,Shahamat2020,Pereira2018,Barata2021} for examples. While grounding of feature visualizations in the relevant clinical task is a commendable effort, the methods to generate the overlaid information have been criticized in regards to their fidelity and specificity~\cite{rudin2019stop,adebayo2018sanity}. Further, as was the case for methods that evaluate the fidelity of transparency information, these methods do not inherently account for human factors.

Lastly, transparent systems can be directly evaluated through user studies on the target population, in which the end users interact with the developed \ac{ML} system to complete a task based on a specific context. In \cite{Folke2021}, the evaluation was centered on the utility of example-based and feature-based explanations for radiologists (8 study participants) to understand the \ac{AI} decision process. Users' understanding was evaluated as the accuracy to predict the \ac{AI}'s diagnosis for a target image and a binary judgement on whether they certify the AI for similar images (and justify using multiple-choice options). Users' agreement with the AI's predictions was measured as well. The empirical evidence suggested that explanations enabled radiologists to develop appropriate trust by making an accurate prediction and judgement of the AI's recommendations. Even though radiologists could complete the task by themselves, a comparison with the team performance was not included, nor the performance of the \ac{AI} model in standalone operation. An alternative evaluation of example-based explanation usefulness was performed in \cite{Sabol2020}, in which pathologists (14 study participants) determined the acceptability of a decision support tool by rating adjectives related to their perceived objectivity, details, reliability, and quality of the system. Compared to a \ac{CNN} without explanations, the subjective ratings were more positive towards the explainable systems. However, neither the team (expert + AI) nor expert baseline performance was evaluated. 
The benefit of involving a dermatologist to complete an image grouping task was demonstrated in \cite{Guo2016}, in which domain knowledge was used to constrain updates of the algorithm's training, resulting in a better grouping performance than a fully automated method. The user evaluation only measured the task performance. These studies that explicitly involve target users to identify whether the envisioned human factors engineering goals were met stand out from the large body of work that did not consider empirical user tests. It is, however, noteworthy that even these exemplary studies are based on very small sample sizes that may not be sufficiently representative of the target users. Careful planning of the study design (including hypothesis statement, experimental design and procedure, participants, and measures) that allows to properly evaluate whether the system achieves the intended goals by adding transparency to the \ac{ML} system is fundamental, especially considering the resources needed and challenges involved in conducting user testing in the healthcare domain.

Even though there were articles that assessed human factors-related properties of the transparency mechanism, a striking majority of articles did not report metrics beyond performance in the main task (n=49) or did not discuss the transparency component at all (n=9). 
Task performance was evaluated in the majority of the articles, 91\% (n=62), and most of them contrasted the performance of the transparent systems with a non-transparent baseline (n=41). Of those, 36 works (88\%) reported improved performance and 5 (12\%) comparable results.

\subsection*{PR: Priors}
\label{sec:Priors}
We differentiate two types of priors that can be used as a source of inspiration to devise transparent \ac{ML} techniques: 1) Priors based on documented knowledge, and especially clinical guidelines considering the unvaried end user specification identified above; and 2) Priors based on computer vision concepts. Most (93\%) articles that incorporated clinical knowledge priors (n=28) directly implemented these priors into the model structure and/or inference, while only 68\% articles with computer vision priors (n=40) provided transparency by the model itself and/or the inference procedure.

A direct way to include clinical knowledge priors was through the prediction, extraction, or use of human-understandable features. Morphological features, \eg, texture, shape and edge features were frequently considered and used to support the transparency of \ac{ML} systems~\cite{Diao2021,Giannini2016,Li2020,Loveymi2020,Kim2018a,Kunapuli2018,Shen2020,Puyol-Anton2020a}. Biomarkers for specific problems, \eg, \ac{EDV} in cardiac MRI~\cite{Puyol-Anton2020,Wongvibulsin2020} and mean diameter, consistency, and margin of pulmonary nodules~\cite{Lin2020} were commonly computed to establish transparency. For problems with a well-established image reporting and diagnosis systems, routinely-used clinical features, \eg, \ac{LI-RADS} features for \ac{HCC} classification~\cite{Wang2019a} or \ac{BI-RADS} for breast mass~\cite{Kim2018} suggested that the \ac{ML} systems may be intuitively interpretable to experts that are already familiar with these guidelines. Human-understandable features relevant to the task domain were extracted from pathology images, \eg, area and tissue structure features~\cite{Diao2021}. Radiomic features were also computed to establish the transparency of \ac{ML} systems~\cite{Kunapuli2018,Yeche2019}.

Besides human-understandable features, clinical knowledge can be used to guide the incorporation of transparency within a model. 
Some articles (n=11) mimicked or started from clinical guidelines and workflows to construct the \ac{ML} systems \cite{Liu2019a,Liu2021,MacCormick2019,Kim2018a,Kim2018,Shahamat2020,Oktay2018,Ren2021,Velikova2013,Zhu2019}. \cite{Liu2019a,Liu2021,MacCormick2019} followed the clinical workflow to encode multiple sources of images and fused the encoded information for the final prediction. Other works followed the specific clinical guidelines of the problems to create transparent systems. \cite{Shahamat2020} split brain \acp{MRI} into 96 clinical meaningful regions as would be done in established clinical workflows and analyze all the regions separately. Some other clinical knowledge priors were also presented. \cite{Codella2018,Barata2021,Yan2019,chen2019deep} established a hierarchical label structure according to clinical taxonomy for image classification. \cite{Dong2021} leveraged the transparency from the correlation between the changes of polarization characteristics and the pathological development of cervical precancerous lesions. Clinical knowledge from human experts was used to refine an image grouping algorithm through an interactive mechanism in which experts iteratively provided inputs to the model~\cite{Guo2016}. %allowed interactive human expert inputs to introduce clinical knowledge iteratively for image grouping.

Priors that were derived from computer vision concepts rather than the clinical workflow were usually not specific or limited to a single application. The justification of transparency with computer vision priors was more general than that with clinical knowledge priors. Image visualization-based techniques to achieve transparency were most commonly considered in image classification problems. Common ways of retrieving relevance information were: Visual relevancy through attention~\cite{AbdelMagid2020,Afshar2021,Fan2020,Graziani2020,An2021,He2021,Hou2019,Huang2019,Morvan2020,Saleem2021,Singla2018,Sun2020,Xu2021}; region occlusion by blank areas~\cite{DeSousa2020,Quellec2021} or healthy-looking regions~\cite{Uzunova2019}; and other techniques such as supervision of activated image regions by clinically relevant areas~\cite{Liao2020,Shinde2019,Khaleel2021,Pereira2018,Verma2019,Wang2019,Zhao2018}, and image similarity~\cite{Folke2021}. Feature-based computer vision transparency priors focused on the impact of feature evolution or perturbation on the decoded output. Encoded features were evolved according to the gradient ascent direction to create the evolution of the decoded image from one class to the other \cite{Couteaux2019,Biffi2018,Silva2018}. Some articles directly analyzed the feature sensitivity to the final prediction by feature perturbation \cite{Couteaux2019,Li2018,Venugopalan2021} and importance analysis \cite{Hao2020,Li2021,Pirovano2020}, feature distribution \cite{Sari2019,Venugopalan2021} or image distribution based on encoded features \cite{Janik2021,Peng2019}. Confidence calibration and uncertainty estimation also increased the transparency of the \ac{ML} systems \cite{Carneiro2020,Sabol2020,Tanno2021}. 

Even though we attempted to identify the type of prior evidence used to justify the development of a specific algorithm in each \ac{ML} system, none of the included articles formally described the process to formulate such priors to achieve transparency in the proposed system. While the use of clinical guidelines and routine workflows may provide Level 2 evidence in support of the method affording transparency if the end users are matched with those priors, relying solely on computer vision techniques may not provide the same level of justification. This is because computer vision algorithms are often developed as an analysis tool for \ac{ML} developers to verify model correctness, but are not primarily designed nor evaluated for use in end user-centered interfaces. The lack of justification and formal processes to inform design choices at the early stages of model development results in substantial risk of creating transparent systems that rely on inaccurate, unintelligible, or irrelevant insights for end users. Being explicit about the assumptions and evidence available in support of the envisioned transparent \ac{ML} system is paramount to build fewer but better-justified transparent \ac{ML} systems that are more likely to live up to expectations in final user testing, the resources for which are heavily constrained.

\subsection*{T: Task}
\label{sec:Task}
Various types of medical image analysis tasks were explored in the included articles. Most of the articles (n=57) proposed transparent \ac{ML} algorithms for classification and detection problems, such as image classification and abnormality detection. Three-dimensional (3D) radiology images (n=24) and pathological images (n=15) were the most popular modalities involved in the development of transparent algorithms. The complex nature of both 3D imaging in radiology and pathological images makes image analysis tasks more time consuming than 2D image analysis that is more prevalent in other specialities, such as dermatology, which motivates transparency as an alternative to complete human image analysis to save time while retaining trustworthiness.  
In detail, classification problems in 3D radiological images and pathological images included abnormality detection in \ac{CT} scans~\cite{Afshar2021,An2021,Loveymi2020,Kunapuli2018,Singla2018,Yan2019,Zhao2018,Zhu2019}, \acp{MRI}~\cite{Graziani2020,Biffi2018,He2021,Li2018,Liu2019,Shinde2019,Shahamat2020,Li2021,Uzunova2019,Puyol-Anton2020a,Puyol-Anton2020,Venugopalan2021,Wang2019a,Wongvibulsin2020}, pathology images~\cite{AbdelMagid2020,DeSousa2020,Diao2021,Dong2021,Fan2020,Graziani2020,Hao2020,Huang2019,Sabol2020,Sari2019,Li2021,Peng2019,Pirovano2020,Yang2020,An2021} and \ac{PET} images~\cite{Morvan2020}. Mammography dominated the 2D radiology image applications~\cite{Liu2021,Silva2018,Kim2018a,Kim2018,Shen2020,Velikova2013,Verma2019,Wang2019,Yeche2019}, mainly focusing on breast cancer classification and mass detection. For other 2D radiology image applications,~\cite{Folke2021,Ren2021} aimed at pneumonia and pneumothorax prediction from chest X-rays and~\cite{Liu2019a} created a transparent model for liver fibrosis stage prediction in liver ultrasound images. Classification and detection tasks were explored in other clinical specialities, including melanoma~\cite{Codella2018} and skin lesion grade prediction~\cite{Graziani2020,Barata2021,Silva2018} in dermatology, glaucoma detection from fundus images~\cite{Liao2020,MacCormick2019,Xu2021} and retinopathy diagnosis~\cite{Quellec2021} in ophthalmology, and polyp classification from colonoscopy images in gastroenterology~\cite{Carneiro2020,Khaleel2021}.

Segmentation was another major application field (n=9). Research about transparency mainly focused on segmentation problems for brain and cardiac \acp{MRI}~\cite{Giannini2016,Hou2019,Janik2021,Saleem2021,Oktay2018,Pereira2018,Sun2020}. Other segmentation problems included mass segmentation in mammograms~\cite{Shen2020}, cardiac segmentation in ultrasound~\cite{Oktay2018}, liver tumor segmentation in hepatic \ac{CT} images, and skin lesion segmentation in dermatological images~\cite{Graziani2020}. There also existed other applications, \eg, image grouping in dermatological images~\cite{Guo2016} and image enhancement (super resolution task) in brain \acp{MRI}~\cite{Tanno2021} and cardiac \acp{MRI}~\cite{Oktay2018}.

Most of the application tasks were routinely performed by human experts in current clinical practice (n=60). A much smaller sample of articles (n=4) aimed to build transparent systems for much more difficult tasks where no human baseline exists, \eg, 5-class molecular phenotype classification from \acp{WSI}~\cite{Diao2021,Khaleel2021}, 5-class polyp classification from colonoscopy images~\cite{Carneiro2020}, cardiac resynchronization therapy response prediction from cardiac \acp{MRI}~\cite{Puyol-Anton2020a}, and super resolution of brain \acp{MRI}~\cite{Tanno2021}. The remaining articles (n=4) did not include explicit information on whether human baselines and established criteria exist for the envisioned application, \eg, magnification level and nuclei area prediction in breast cancer histology images~\cite{Graziani2020}, age estimation in brain \acp{MRI}~\cite{He2021}, \ac{AD} status in \acp{DTI}, and risk of sudden cardiac death prediction in cardiac \acp{MRI}~\cite{Wongvibulsin2020}.
As previously mentioned, tasks that are routinely performed in clinical evidence may have robust human baselines and clinical guidelines to guide transparent \ac{ML} development. Applications that are beyond the current possibilities, however, require a more nuanced and human-centered approach that should involve the target end users as early as possible to verify that the assumptions that drive transparency are valid.

\section*{Data Availability}
Figure~2 contains images from the ORIGA~\cite{zhang2010origa} and BraTS2020 datasets~\cite{menze2014multimodal}. The ORIGA dataset is a public dataset at Kaggle website (\url{https://www.kaggle.com/datasets/sshikamaru/glaucoma-detection/metadata}). The BraTS2020 dataset is also a public dataset at Kaggle website (\url{https://www.kaggle.com/datasets/awsaf49/brats2020-training-data}).

\section*{Acknowledgements}
We acknowledge all authors for the contribution. No funding is included.

\section*{Author contributions}
All authors contributed to the conception and design of the study. HC and CG contributed to the literature search and data extraction. HC and CG contributed to data analysis and interpretation. All authors contributed to writing the manuscript, and all authors approved the manuscript. All authors guaranteed the integrity of the work. HC and CG contributed equally in this work and are co-first authors.

\section*{Competing interests}
No conflict of interests exist.

\bibliographystyle{naturemag}
\bibliography{reference}

\end{document}

\section{The INTRPRT guideline: A case study}
\label{app:case}
To contextualize the \emph{INTRPRT guideline}, we present case studies to demonstrate the envisioned use. Because none of the surveyed papers take into account all aspects mentioned in the \emph{INTRPRT guideline}, we include three published papers that are representative of different parts of the \emph{INTRPRT guideline}. Case 1 (C1): Xie et al.~\cite{xie2020chexplain} focused on the formative user research stage to determine physicians' needs when exploring and understanding an \ac{AI}-generated analysis report in the context of chest X-rays diagnosis. Through surveys, a co-design process, and user evaluations, different features of the explanations that physicians expect when they interact with the system were identified. Case 2 (C2): Motivated by the fact that the radiologists characterize breast masses according to \ac{BI-RADS} (public evidence), Kim et al.~\cite{Kim2018} directly encoded the \ac{BI-RADS} characteristics of breast masses to build a deep learning model for mass classification. Case 3 (C3): Sabol et al.~\cite{Sabol2020} created an explainable system for colorectal cancer diagnosis from histopathological images by providing human-friendly explanations for a certain prediction. The validation of the system included the assessment of transparency through a human factors evaluation with 14 pathologists who interacted with the system through a graphical user interface.

We identify which aspects within the three cases follow the \emph{INTRPRT guideline}: 

\emph{G1: Specify the clinical scenario, constraints, requirements, and end users.}\\
In C1, \ac{ML} designers determined the requirements in current clinical practice by conducting paired surveys on physicians and radiologists, where the latter provide a diagnosis and the former have to interpret the results. The target users (referring physicians) were identified within the context in which the system will be used. 

\emph{G2: Justify the choice of transparency and determine the level of evidence.}\\
In C1, to inform the design of \ac{AI}-enabled chest X-ray analysis, iteratively developed evidence was generated by involving potential end users of the system through interactions with low and high-fidelity prototypes. The design choices to build the prototypes were based on the initial needs identified with a survey of explanations between referring physicians and radiologists. \\
In C2, the ac{BI-RADS} standard serves as public evidence because it is widely accepted and applied among clinicians for mass classification.

\emph{G3: Clarify how the model follows the justification of transparency.}\\
In C2, the public evidence, namely \ac{BI-RADS} characteristics of breast masses, are encoded as features and concatenated with deep encoded image features for the final tissue classification. As a result, \ac{BI-RADS} characteristics are explicitly extracted and have direct impact in the final decision making.

\emph{G4: Determine how to communicate with end users.}\\
In C1, a graphical interface was implemented as a high-fidelity prototype. The interface presented one patient's case at a time, displaying the \ac{CXR} image, the significant observations generated from an \ac{AI} model (as textual labels), and eight explanations features that were manually generated, such as highlighting evidence towards a specific diagnosis in the image, probabilities for each possible conclusion or comparisons with previous patients' cases.\\
C3 created a graphical user interface that showed the original image of the \ac{WSI} and the corresponding label map with a color code for different tissue types. Pathologists can examine an arbitrary area of the \ac{WSI} by clicking on the desired area. Subsequently, the outcomes of the system were displayed, including the prediction result with a semantical explanation. Besides, a visualization of the training image most similar to the current one as well as training images with other tissue types are presented as references and support for the prediction.

\emph{G5: Report task performance of the ML systems.}\\
In C2, quantitative results of the method with/without implementing \ac{BI-RADS} characteristics are presented on a public mammogram database. Both accuracy and \ac{AUC} are used as evaluation metrics of the model itself. \\
C3 reported the accuracy of the classifier in a class-balanced dataset of tissue slides. 

\emph{G6: Assess correctness and human factors of system transparency. } \\
C3 involved 14 pathologists to evaluate four human factors of the system: usefulness, level of detail, reliability, and experience quality. Pathologists were first asked to examine 20 arbitrary areas in a graphical user interface and evaluate the prediction outcome. At the end of the experiment session, every participant was asked to fill out a questionnaire based on their perception of the system. 

Lastly, we present examples on where and how authors in the three cases could integrate the use of the \emph{INTRPRT guideline} in their articles: 

C1: At the beginning of the Method section: ``We were primarily concerned on the user-centered iterative design of a proof-of-concept system prototype of an AI-based medical image analysis tool with different explanation features to assist physicians. The outcome of the user-centered empirical formative research approach includes recommendations and insights that may benefit future development of explainable algorithms for medical image analysis.''

C2: In the Method section: ``We followed the widely accepted \ac{BI-RADS} standard for breast mass assessment as a readily interpretable backbone as a basis for our deep network.'' In the Results section: ``The verification of the system included a task performance comparison against a model without the interpretable \ac{BI-RADS} component on a public database. Additional qualitative visualizations showed relevant areas where the model exploited more information.'' 

C3: At the end of the Introduction section: ``We developed and discussed the mathematical structure of a classifier that provides different outputs to explain the plausibility of the decision. A quantitative evaluation of the system's performance was conducted on a public database. We also report the system's acceptability assessed through a user study with 14 pathologists.''

\section{Search strategy}
\label{app:search_strategy}
We use the following search term for PubMed and EMBASE to screen titles, abstracts, and keywords of all available records:

(``interpretable'' OR ``explainable'' OR ``interpretability'' OR ``explainability'' OR ``interpretation'' OR ``explanation'' OR ``interpreting'' OR ``explaining'' OR ``interpret'' OR ``explain'') AND (``artificial intelligence'' OR ``deep learning'' OR ``machine learning'' OR ``neural network'') AND (``image'' OR ``imaging'') AND (``healthcare'' OR ``health care'' OR ``clinical'' OR ``medical'')

In addition to the above search terms, Compendex offers ``controlled terms'' to better locate desired articles. We first filter all records with the following search term for titles, abstracts, and keywords:

(interpret* OR explain* OR ``explanation'')

Then we use ``controlled terms'' to further filter all the remaining records:

(``artificial intelligence'' OR ``neural networks'' OR ``machine learning'' OR ``deep learning'' OR ``deep neural networks'' OR ``convolutional neural networks'' OR ``learning systems'' OR ``supervised learning'' OR ``network architecture'') AND (``medical imaging'' OR ``medical image processing'' OR ``medical computing'' OR ``tissue'')

For screening in all three databases, we also exclude articles with (``survey'' OR ``review'') in the tile and (``workshop'') in the title and abstract.

\section{Details of screening and full-text review}
\label{app:screen}
The initial search resulted in $2508$ records, and after removal of duplicates, $1731$ unique studies were included for screening. During screening, $1514$ articles were excluded because they 1) were not transparent methods (n=947); 2) were not imaging \ac{ML} methods (n=422); 3) only had simple visualization explanations (n=129) and 4) were not for medical problems (n=19). We found unsubstantiated claims around transparency with only simple visualizations such as \acp{CAM} widely occur in \ac{ML} methods or using existing transparent methods. As a result, we excluded them and focused on articles with other transparent \ac{ML} methods. The remaining $217$ articles were included in full-text review. In total, $149$ records were further excluded because they 1) used exactly the same transparency mechanism as previously proposed work for natural image problems (n=48); 2) were not transparent methods (n=43); 3) were not long articles and therefore could not be analyzed with the required detail (n=24); 4) were not imaging \ac{ML} methods (n=21); 5) were not for medical problems (n=5); 6) did not have available full text (n=5); 7) were repeated works (n=3). The criterion for long articles we applied was single-column articles longer or equal to 8 pages or double-column articles longer or equal to 6 pages, excluding reference pages.

\section{Data extraction}
\label{app:extraction}
Supplementary Table~\ref{tab:extraction} describes in detail our data extraction approach for the studies included in this review. Supplementary Tables~\ref{tab:extraction_table1}~and~\ref{tab:extraction_table2} present the details of each study included in the review in the design preparation and implementation, respectively.

\begin{table}[H]
\centering
\caption{Data extraction strategy related to the six themes.}
\begin{center}
\begin{tabular}{ |>{\centering\arraybackslash}p{0.18\textwidth}|p{0.3\textwidth}|p{0.4\textwidth}| } 
\hline
\textbf{Theme} & \textbf{Item} & \textbf{Description}\\
\hline
\multirow{2}{*}{Incorporation} & Clinician engineering team & Any clinical stakeholders part of the study team and author list\\
& Formative user research & Any technique to understand the target population\\
\hline
Target & End users & Users of the systems\\
\hline
\multirow{2}{*}{Prior} & Justification of transparency & Description of the choice of transparency\\
                       & Prior type & Computer vision / clinical knowledge prior\\
\hline
\multirow{2}{*}{Task} & Inputs \& outputs & Task inputs \& outputs\\
                      & Task difficulty & Routine / super-human tasks \\
\hline
\multirow{2}{*}{Interpretability} & Technical mechanism & Explicit transparency technique\\
& Transparency type & Interpretable (provides its own explanations) / Explainable (needs post-hoc explanations)\\
\hline
\multirow{5}{*}{Reporting} & Metrics & Task performance evaluation and transparency assessment metrics\\
                            & Transparency performance & Performance against comparable baseline models\\
                            & Incorporation performance & Performance of human-AI incorporation against AI alone \\
%                             & Human factors & Assessed human factors\\
                            & Human subjects & Number of end users involved in evaluation\\
\hline
\end{tabular}
\end{center}
\label{tab:extraction}
\end{table}

\footnotesize
\begin{xltabular}{\textwidth}{|>{\centering\arraybackslash}m{0.05\textwidth}|>{\centering\arraybackslash}m{0.07\textwidth}|>{\centering\arraybackslash}m{0.12\textwidth}|>{\centering\arraybackslash}m{0.11\textwidth}|>{\centering\arraybackslash}m{0.09\textwidth}|>{\centering\arraybackslash}m{0.09\textwidth}|>{\centering\arraybackslash}m{0.12\textwidth}|>{\centering\arraybackslash}m{0.10\textwidth}|}
\caption{A summary of transparency design preparation details of each study reviewed. The definition of each column is the same as in Supplementary Table~\ref{tab:extraction}.}
\label{tab:extraction_table1}\\
\hline
\textbf{Study ID}&\textbf{Author team }& \textbf{End users}&\textbf{Task type}&\textbf{Formative user research}&\textbf{Task difficulty}&\textbf{Clear justification of transparency}&\textbf{Prior type} \\
\hline
\citep{AbdelMagid2020} & Yes & Not specified & Prediction & None & Routine & Yes & Computer vision\\
\hline
\citep{Afshar2021} & Yes & Decision providers & Prediction & None & Routine & No & Computer vision\\
\hline
\citep{Codella2018} & Yes & Decision providers & Prediction & None & Routine & No & Clinical knowledge\\
\hline
\citep{Couteaux2019} & No & Not specified & Segmentation & None & Routine & Yes & Computer vision\\
\hline
\citep{DeSousa2020} & No & Not specified & Prediction & None & Routine & Yes & Computer vision\\
\hline
\citep{Diao2021} & Yes & Decision providers & Prediction & None & Super-human & Yes & Clinical knowledge\\
\hline
\citep{Dong2021} & Yes & Decision providers & Prediction & None & Routine & Yes & Clinical knowledge\\
\hline
\citep{Fan2020} & Yes & Not specified & Prediction & None & Routine & No & Computer vision\\
\hline
\citep{Folke2021} & No & Decision providers & Prediction & None & Routine & Yes & Computer vision\\
\hline
\citep{Giannini2016} & Yes & Decision providers & Segmentation & None & Routine & No & Clinical knowledge\\
\hline
\citep{Graziani2020} & Cannot tell & Not specified & Prediction & None & Cannot tell & Yes & Computer vision\\
\hline
\citep{Graziani2020} & Yes & Not specified & Segmentation& None & Routine & Yes & Computer vision\\
\hline
\citep{Guo2016} & Yes & Decision providers & Image grouping & None & Routine & Yes & Clinical knowledge\\
\hline
\citep{An2021} & No & Not specified & Prediction & None & Routine & Yes & Computer vision\\
\hline
\citep{Barata2021} & No & Decision providers & Prediction & None & Routine & Yes & Clinical knowledge\\
\hline
\citep{Biffi2018} & Yes & Not specified & Prediction & None & Routine & No & Computer vision\\
\hline
\citep{Carneiro2020} & Yes & Decision providers & Prediction & None & Super-human & Yes & Computer vision\\
\hline
\citep{Hao2020} & Yes & Decision providers & Prediction & None & Routine & Yes & Computer vision\\
\hline
\citep{He2021} & Yes & Not specified & prediction & None & Cannot tell & Yes & Computer vision\\
\hline
\citep{Hou2019} & Yes & Not specified & Segmentation & None & Routine & Yes & Computer vision\\
\hline
\citep{Huang2019} & No & Decision providers & Prediction & None & Routine & Yes & Computer vision\\
\hline
\citep{Li2018} & Yes & Not specified & Prediction  & None & Routine & Yes & Computer vision\\
\hline
\citep{Li2020} & Yes & Decision providers & Prediction & None & Routine & Yes & Clinical knowledge\\
\hline
\citep{Liao2020} & No & Not specified & Prediction  & None & Routine & Yes & Computer vision\\
\hline
\citep{Lin2020} & Yes & Decision providers & Prediction & None & Routine & Yes & Clinical knowledge\\
\hline
\citep{Liu2019a} & Cannot tell & Not specified & Prediction & None & Routine & Yes & Clinical knowledge\\
\hline
\citep{Liu2019} & No & Not specified & Prediction & None & Cannot tell & Yes & Clinical knowledge\\
\hline
\citep{Liu2021} & No & Decision providers & Prediction & None & Routine & Yes & Clinical knowledge\\
\hline
\citep{Loveymi2020} & No & Not specified & Predictions & None & Routine & No & Clinical knowledge\\
\hline
\citep{MacCormick2019} & Yes & Not specified & Prediction & None & Routine & Yes & Clinical knowledge\\
\hline
\citep{Morvan2020} & Yes & Not specified & Prediction & None & Routine & Yes & Computer vision\\
\hline
\citep{Shinde2019} & Yes & Not specified & Prediction & None & Routine & Yes & Computer vision\\
\hline
\citep{Silva2018} & Yes & Not specified & Prediction & None & Routine & Yes & Computer vision\\
\hline
\citep{Janik2021} & Yes & Decision providers & Segmentation & None & Routine & Yes & Computer vision\\
\hline
\citep{Khaleel2021} & Yes & Not specified & Prediction & None & Super-human & Yes & Computer vision\\
\hline
\citep{Kim2018a} & No & Decision providers & Prediction & None & Routine & Yes & Clinical knowledge\\
\hline
\citep{Kim2018} & No & Decision providers & Prediction & None & Routine & Yes & Clinical knowledge\\
\hline
\citep{Kunapuli2018} & Yes & Not specified & prediction & None & Routine & Yes & Clinical knowledge\\
\hline
\citep{Sabol2020} & Yes & Decision providers & Prediction & None & Routine & Yes & Computer vision\\
\hline
\citep{Saleem2021} & No & Not specified & Segmentation & None & Routine & No & Computer vision\\
\hline
\citep{Sari2019} & No & Not specified & prediction & None & Routine & No & Computer vision\\
\hline
\citep{Shahamat2020} & No & Not specified & Prediction & None & Routine & Yes & Clinical knowledge\\
\hline
\citep{Shen2020} & No & Not specified & Prediction & None & Routine & Yes & Clinical knowledge\\
\hline
\citep{Li2021} & No & Not specified & prediction & None & Routine & Yes & Computer vision\\
\hline
\citep{Oktay2018} & Yes & Not specified & Segmentation  & None & Routine & Yes & Clinical knowledge\\
\hline
\citep{Peng2019} & Yes & Decision providers & Prediction & None & Routine & Yes & Computer vision\\
\hline
\citep{Pereira2018} & Yes & Not specified & Segmentation & None & Routine & Yes & Computer vision\\
\hline
\citep{Uzunova2019} & No & Decision providers & prediction & None & Routine & Yes & Computer vision\\
\hline
\citep{Pirovano2020} & No & Decision providers & Prediction & None & Routine & No & Computer vision\\
\hline
\citep{Puyol-Anton2020a} & No & Decision providers & Prediction & None & Super-human & Yes & Clinical knowledge\\
\hline
\citep{Puyol-Anton2020} & Yes & Decision providers & Prediction & None & Routine & Yes & Clinical knowledge\\
\hline
\citep{Quellec2021} & Yes & Decision providers & Prediction & None & Routine & No & Computer vision\\
\hline
\citep{Ren2021} & Yes & Decision providers & prediction & None & Routine & No & Clinical knowledge\\
\hline
\citep{Singla2018} & Yes & Not specified & Prediction & None & Routine & Yes & Computer vision\\
\hline
\citep{Sun2020} & Yes & Not specified & Segmentation & None & Routine & Yes & Computer vision\\
\hline
\citep{Tanno2021} & Yes & Decision providers & Super-resolution & None & Super-human & Yes & Computer vision\\
\hline
\citep{Velikova2013} & Yes & Not specified & Prediction & None & Routine & Yes & Clinical knowledge\\
\hline
\citep{Venugopalan2021} & No & Decision providers & Prediction & None & Routine & Yes & Computer vision\\
\hline
\citep{Verma2019} & No & Decision providers & prediction & None & Routine & Yes & Computer vision\\
\hline
\citep{Wang2019a} & Yes & Decision providers & Prediction & None & Routine & No & Clinical knowledge\\
\hline
\citep{Wang2019} & Yes & Decision providers & Prediction & None & Routine & No & Computer vision\\
\hline
\citep{Wongvibulsin2020} & Yes & Decision providers & Prediction & None & Cannot tell & No & Clinical knowledge\\
\hline
\citep{Xu2021} & Yes & Not specified & Prediction & None & Routine & Yes & Computer vision\\
\hline
\citep{Yan2019} & Yes & Decision providers & Prediction & None & Routine & Yes & Clinical knowledge\\
\hline
\citep{Yang2020} & Yes & Decision providers & Prediction & None & Routine & No & Computer vision\\
\hline
\citep{Yeche2019} & No & Decision providers & Prediction & None & Routine & Yes & Clinical knowledge\\
\hline
\citep{Zhao2018} & No & Not specified & Prediction & None & Routine & Yes & Computer vision\\
\hline
\citep{Zhu2019} & No & Not specified & Prediction & None & Routine & Yes & Clinical knowledge\\
\hline 
\end{xltabular}

\begin{xltabular}{\textwidth}{|>{\centering\arraybackslash}m{0.05\textwidth}|>{\centering\arraybackslash}m{0.13\textwidth}|>{\centering\arraybackslash}m{0.14\textwidth}|>{\centering\arraybackslash}m{0.13\textwidth}|>{\centering\arraybackslash}m{0.13\textwidth}|>{\centering\arraybackslash}m{0.13\textwidth}|>{\centering\arraybackslash}m{0.08\textwidth}|}

\caption{A summary of transparency design implementation and validation details of each study reviewed. The definition of each column is the same as in Supplementary Table~\ref{tab:extraction}.}\label{tab:extraction_table2}\\

\hline
\textbf{Study ID}&\textbf{Transparency type}& \textbf{Technical mechanism}&\textbf{Transparency assessment}&\textbf{Transparency performance}&\textbf{Incorporation performance}&\textbf{Human subjects}\\
\hline
\citep{AbdelMagid2020} & Interpretable & Attention & Spearman coefficients & Better & None & None\\
\hline
\citep{Afshar2021} & Interpretable & Attention & None & Better & None & None\\
\hline
\citep{Codella2018} & Interpretable & Clustering & None & Better & None & None\\
\hline
\citep{Couteaux2019} & Explainable & activation maximization & None & None & None & None\\
\hline
\citep{DeSousa2020} & Explainable & Input perturbation & Visualization & None & None & None\\
\hline
\citep{Diao2021} & Interpretable & Automatic extraction of clinically important features & Similarity & None & None & None\\
\hline
\citep{Dong2021} & Interpretable & Automatic extraction of clinically important features & Visualization & None & None & None\\
\hline
\citep{Fan2020} & Interpretable & Attention & None & Better & None & None\\
\hline
\citep{Folke2021} & Explainable & Saliency maps & Level of trust & None & None & 8\\
\hline
\citep{Giannini2016} & Interpretable & Hand-crafted feature computation & Confusion matrix & None & None & None\\
\hline
\citep{Graziani2020} & Interpretable & Network architecture pruning & Kappa, $R^2$ matrix & Better & None & None\\
\hline
\citep{Gu2021} & Interpretable & Attention & Visualization & Better & None & None\\
\hline
\citep{Guo2016} & Interpretable & Clustering & None & Better & Better & 1\\
\hline
\citep{An2021} & Interpretable & Attention & None & None & None & None\\
\hline
\citep{Barata2021} & Interpretable & Decoded by simple transparent models & None & Comparable & None & None\\
\hline
\citep{Biffi2018} & Interpretable & Latent variable evolution & None & None & None & None\\
\hline
\citep{Carneiro2020} & Interpretable & Uncertainty estimation / confidence calibration & None & Better & None & None\\
\hline
\citep{Hao2020} & Interpretable & Ranking of features & C-index, visualization & Better & None & None\\
\hline
\citep{He2021} & Interpretable & Attention & Pearson correlation, visualization & Better & None & None\\
\hline
\citep{Hou2019} & Interpretable & Attention & None & Better & None & None\\
\hline
\citep{Huang2019} & Interpretable & Attention & None & Better & None & None\\
\hline
\citep{Li2018} & Explainable & Input corruption & Visualization & None & None & None\\
\hline
\citep{Li2020} & Interpretable & Decoded by simple transparent models & T-SNE visualization & Better & None & None\\
\hline
\citep{Liao2020} & Explainable & Visualization & Visualization & Better & None & None\\
\hline
\citep{Lin2020} & Interpretable & Automatic extraction of clinically important features & None & None & None & None\\
\hline
\citep{Liu2019a} & Interpretable & Network structure modification & None & Better & None & None\\
\hline
\citep{Liu2019} & Interpretable & Causal inference & Causal relationships, visualization & None & None & None\\
\hline
\citep{Liu2021} & Interpretable & Network structure modification & Visualization & Better & None & None\\
\hline
\citep{Loveymi2020} & Interpretable & Automatic extraction of clinically important features & None & Better & None & None\\
\hline
\citep{MacCormick2019} & Interpretable & Automatic extraction of clinically important features & None & None & None & None\\
\hline
\citep{Morvan2020} & Interpretable & Attention & C-index & Better & None & None\\
\hline
\citep{Shinde2019} & Explainable & Class activation mapping & Visualization & Better & None & None\\
\hline
\citep{Silva2018} & Explainable & Decoded by simple transparent models & Correctness, completeness and compactness & None & None & None\\
\hline
\citep{Janik2021} & Interpretable & Clustering & None & None & None & None\\
\hline
\citep{Khaleel2021} & Interpretable & Decoded by simple transparent models & Faithfulness, relevance scores & None & None & None\\
\hline
\citep{Kim2018a} & Interpretable & Hand-crafted feature computation & None & Better & None & None\\
\hline
\citep{Kim2018} & Interpretable & Automatic extraction of clinically important features & Visualization & Better & None & None\\
\hline
\citep{Kunapuli2018} & Interpretable & Relation analysis & None & Better & None & None\\
\hline
\citep{Sabol2020} & Interpretable & Uncertainty estimation / confidence calibration & Certainty rate \& error, User perception & Comparable & None & 14\\
\hline
\citep{Saleem2021} & Explainable & Attention & Deletion metric & None & None & None\\
\hline
\citep{Sari2019} & Interpretable & Clustering & Visualization & Better & None & None\\
\hline
\citep{Shahamat2020} & Interpretable & Attention with domain knowledge & None & Better & None & None\\
\hline
\citep{Shen2020} & Interpretable & Decoded by simple transparent models  & None & Better & None & None\\
\hline
\citep{Li2021} & Interpretable & Representative feature extraction & None & Better & None & None\\
\hline
\citep{Oktay2018} & Interpretable & Prior knowledge into latent space  & Agreement & Better & None & 2\\
\hline
\citep{Peng2019} & Interpretable & Image retrieval &Visualization & Better & None & None\\
\hline
\citep{Pereira2018} & Explainable & Decoded by simple transparent models & Visualization & Comparable & None & None\\
\hline
\citep{Uzunova2019} & Interpretable & Perturbation analysis & Visualization & Better & None & None\\
\hline
\citep{Pirovano2020} & Explainable & Feature importance analysis & Visualizations & None & None & None\\
\hline
\citep{Puyol-Anton2020a} & Interpretable & Network structure modification  & Visualization & Comparable & None & None\\
\hline
\citep{Puyol-Anton2020} & Interpretable & Relation analysis  & $R^2$ matrix & None & None & None\\
\hline
\citep{Quellec2021} & Interpretable & Automatic extraction of clinically important features & Kappa & Better & None & None\\
\hline
\citep{Ren2021} & Interpretable & Causal inference & None & Comparable & None & None\\
\hline
\citep{Singla2018} & Interpretable & Attention & $R^2$ matrix, visualization & Better & None & None\\
\hline
\citep{Sun2020} & Interpretable & Attention & Visualization & Better & None & None\\
\hline
\citep{Tanno2021} & Interpretable & Uncertainty estimation / confidence calibration & Reliability & Better & None & None\\
\hline
\citep{Velikova2013} & Interpretable & Causal inference & None & None & None & None\\
\hline
\citep{Venugopalan2021} & Explainable & Perturbation analysis & None & None & None & None\\
\hline
\citep{Verma2019} & Interpretable & Decoded by simple transparent models & Visualization & None & None & None\\
\hline
\citep{Wang2019a} & Interpretable & Automatic extraction of clinically important features & None & None & None & None\\
\hline
\citep{Wang2019} & Explainable & Visualization & Visualization & None & None & None\\
\hline
\citep{Wongvibulsin2020} & Explainable & Relation analysis & Visualization & None & None & None\\
\hline
\citep{Xu2021} & Interpretable & Attention & Visualization & Better & None & None\\
\hline
\citep{Yan2019} & Interpretable & Decoded by simple transparent models & Visualization & Better & None & None\\
\hline
\citep{Yang2020} & Interpretable & Attention & Visualization & Better & None & None\\
\hline
\citep{Yeche2019} & Interpretable & Concept's importance analysis  & $R^2$ matrix & None & None & None\\
\hline
\citep{Zhao2018} & Explainable & Backpropagation guidance & Visualization, Kendall's Tau metric & None & None & None\\
\hline
\citep{Zhu2019} & Explainable & Perturbation analysis & Shapley values & None & None & None\\
\hline
\end{xltabular}

% \bibliographystyle{vancouver}
% \bibliography{reference}